\documentclass[11pt,a4paper]{article}
\usepackage[utf8]{inputenc}
\usepackage[T1]{fontenc}
\usepackage{lmodern}
\usepackage[margin=1in]{geometry}
\usepackage{amsmath,amssymb,amsthm}
\usepackage{booktabs}
\usepackage{array}
\usepackage{graphicx}
\usepackage{xcolor}
\usepackage{listings}
\usepackage[hyphens]{url}
\usepackage[colorlinks=true,linkcolor=blue!60!black,citecolor=blue!60!black,urlcolor=blue!60!black]{hyperref}

\newsavebox{\mathbox}
\newcommand{\fitmath}[2]{%
  \sbox{\mathbox}{$\displaystyle #2$}%
  \ifdim\wd\mathbox>#1%
    \resizebox{#1}{!}{\usebox{\mathbox}}%
  \else%
    \usebox{\mathbox}%
  \fi%
}

\title{\textbf{Practical Preimages for 19- and 20-Step SHA-256 with Standard Padding}}
\author{Sebastian R. Wain\\\texttt{srw@egont.xyz}\\[0.25em]\small\url{https://github.com/srw/sha256-19-and-20-step-preimages-with-padding}}
\date{}

\begin{document}
\maketitle
\vspace{-1.5em}
\begin{abstract}
We present practical preimage attacks on 19- and 20-step SHA-256 that respect the standard IV, Davies--Meyer feed-forward, and full FIPS 180-4 padding, for arbitrary prescribed 256-bit targets. Building on Basra's algebraic selector saturation and backward padding absorption [9], we introduce three techniques: an exact $O(1)$ inversion of the low padding byte $(W_{13} \bmod 256) = \mathtt{0x80}$ that leaves $(a_5, a_6)$ free for exact context shaping; a causal bit-slice equivalence theorem that prunes $99.74\%$ of right-hand candidate nodes across seven stages; and a closed-form $\mathbb{F}_2$ upper-half inverse paired with compact $Q_0^{-1}$ lookup tables. One CPU core solves arbitrary 19-step padded targets in $1.2\text{-}86.4\text{ s}$ (median $19.5\text{ s}$) after an $18\text{ s}$ table build. A 20-step padded preimage costs $2^{46.6}$ CPU-reference evaluations: $191$ node-hours on one 10-core Apple M4 CPU node (16 threads), or $9.0\text{-}22.2$ GPU-hours on one NVIDIA L4. A companion Zero-$a_3$ bridge inverts the unpadded 20-step compression function in $2^{38.1}$ evaluations. We provide \textbf{46} self-verifying witnesses, including 55-byte preimages of $\text{SHA-256}(\text{``abc''})$ and $0^{256}$. These attacks cover 19--20 of 64 steps and do not threaten full SHA-256.

\end{abstract}

\section{Introduction}

SHA-256 [1] hashes an $L$-bit message $M$ by appending mandatory Merkle--Damgård strengthening (FIPS 180-4 \S{}5.1.1)---a single \texttt{'1'} bit ($\mathtt{0x80}$ at a byte boundary), zero padding bits, and the 64-bit big-endian bit length $L$ in the final two 32-bit words $W_{14}$ and $W_{15}$---before applying a 64-step Davies--Meyer compression function initialized to the standard initial value $\mathrm{IV}$.

In the cryptanalysis of step-reduced SHA-256, a persistent divide has separated \textbf{theoretical} attacks from \textbf{practical (demonstrated)} preimages:
\begin{itemize}
\item \textbf{Theoretical preimage and collision attacks:} Splice-and-cut meet-in-the-middle (MitM), partial-matching, biclique, and MILP-assisted attacks respect FIPS 180-4 padding from 24 steps [2] through 41--43 steps [3, 10], 45 steps [4], and 46--47 steps [8], with related differential and semi-free-start collision advances reaching 24--39 steps for SHA-256 (and 40 steps for SHA-224) [11, 12, 14, 15] building on the generalized characteristic framework introduced for SHA-1 in [13]. Unlike SHA-1, where full-round collisions have been practically computed [17, 18], all 24--47-step SHA-256 preimage attacks require between $2^{240}$ and $2^{255.6}$ compression evaluations and remain strictly theoretical.
\item \textbf{Practical (demonstrated) preimages:} Algebraic and SAT-based encodings produced padded preimages up to 16 steps [5] and three mailing-list challenge messages at 17 steps (posted without an attack specification) [16]. More recently, SAT/Cube-and-Conquer [6, 7] produced verified preimages at 18 and 19 steps for the unpadded compression function (where all 16 message words $W_0, \dots, W_{15}$ are unconstrained 32-bit degrees of freedom), and Basra [9] introduced algebraic selector saturation and context shaping to solve 19- and 20-step unpadded compression targets as well as a 55-byte 20-step padded preimage for a structured target (\texttt{padded\_r20} in [9], dated 17 October 2026).

\end{itemize}
For a single-block 55-byte ($L = 440$-bit) padded message, FIPS 180-4 [1] fixes 72 bits at the end of the message block:
\[
\fitmath{0.86\textwidth}{W_{15} = 440 \ (\mathtt{0x000001b8}), \qquad W_{14} = 0 \ (\mathtt{0x00000000}), \qquad W_{13} \bmod 256 = \mathtt{0x80}.} \tag{E1}
\]
In backward-and-forward inversion of steps $0..18$ (19-step) or $0..19$ (20-step), $W_{13}, W_{14}, W_{15}$ are tightly coupled with the message schedule expansion $W_{16}, \dots, W_{19}$ and the internal state window $(a_5, \dots, a_{12})$. Satisfying all 72 padding bits simultaneously with the selector-saturation and majority-absorption conditions [9] that decouple the left and right halves of the compression function is the central barrier to inverting \textbf{arbitrary externally prescribed 256-bit targets} for the padded 19- and 20-step hash functions.

\subsection{Upfront Demonstration of 19-Step and 20-Step Padded SHA-256 Preimages}

To illustrate our results concretely before entering the algebraic derivations, below are four self-verifying \textbf{FIPS 180-4 padded preimages} for 20-step and 19-step SHA-256 ($\mathrm{SHA256}_{20}$ and $\mathrm{SHA256}_{19}$ with standard $\mathrm{IV}$, Davies--Meyer feed-forward, and mandatory padding [1]; complete 46-witness verification suite in \S{}8 and at \url{https://github.com/srw/sha256-19-and-20-step-preimages-with-padding}):

\begin{enumerate}
\item \textbf{20-Step Single-Block Padded Preimage for FIPS 180-4 Test Vector $H^* = \text{SHA-256}(\text{``abc''})$:}
\begin{lstlisting}
Target H*          = ba7816bf8f01cfea414140de5dae2223b00361a396177a9cb410ff61f20015ad
55-Byte Message M  = edf6e70a7001ae71b900f18119aaa1fdf6c1179b4268891262a9539
                     be04d41236685e0c102e0ad23165e8b7157669819302ddcb3c68226
Padded Block W0..7 = edf6e70a 7001ae71 b900f181 19aaa1fd f6c1179b 42688912 62a9539b e04d4123
Padded Block W8.15 = 6685e0c1 02e0ad23 165e8b71 57669819 302ddcb3 c6822680 00000000 000001b8
Verification       : SHA256_20(M) == H*  (W13 & 0xff = 0x80, W14 = 0, W15 = 440 bits)
\end{lstlisting}
\item \textbf{19-Step Single-Block Padded Preimages for $H^* = \text{SHA-256}(\text{``abc''})$ and $H^* = \mathtt{0000\dots0000}$:}
\begin{lstlisting}
'abc' 55B Message  = 8ca254f8d04dd3fb04efaad6b1867588ffe833de49a144fedbf1839
                     422892dc9f5a9158d8fc063290a14a9a6c55dfa245642a0247ec45a
0^256 55B Message  = fc0c9d8eafb2fa329af316940c42a5eb9773cec39e4f7e396182628
                     7ae767638f92d4ce5fc5eb8cabd7606409364dda27a6c1bb5220246
Verification       : SHA256_19(M_abc) == SHA-256("abc") and SHA256_19(M_zero) == 0^256
\end{lstlisting}
\item \textbf{20-Step Single-Block Padded Preimage for All-Zero Digest $H^* = \mathtt{0000\dots0000}$:}
\begin{lstlisting}
Target H*          = 0000000000000000000000000000000000000000000000000000000000000000
55-Byte Message M  = ec531306c85ad4a4ad211d2e1a03ece4d0ad4a0e6b4c2b9436b3bab
                     dcbd520597196f52c042b8cd95f1a39f71b58b47b0647c78e65416f
Padded Block W8.15 = 7196f52c 042b8cd9 5f1a39f7 1b58b47b 0647c78e 65416f80 00000000 000001b8
Verification       : SHA256_20(M) == 0^256  (W13 & 0xff = 0x80, W14 = 0, W15 = 440 bits)
\end{lstlisting}
\item \textbf{20-Step \& 19-Step Two-Block (119-Byte, Plain-ASCII Block 1) Padded Preimages for $H^* = \text{SHA-256}(\text{``''})$:}
\begin{lstlisting}
Target H*          = e3b0c44298fc1c149afbf4c8996fb92427ae41e4649b934ca495991b7852b855
Block 1 (64B ASCII)= "This first block is plain text; the second one was computed....."
20-Step Block 2 M2 = df28e21bb6b61202ba87baa72096ee0a3ac325cf1f12056e7cfe2ae
                     1735b4d106b8f6e0b22eb9b504ed9368991944b0a115a934f17c2c4
19-Step Block 2 M2 = fe409cc930128a720481f86c0503d7a018519a129df449139720bb5
                     eef520895f8b515685bb404817be301880f0caabe02fc9e5f1c95c0
Padding Suffix     : ...80 00000000 000003b8 (L = 952 bits = 119 bytes; full Merkle-Damgard 2-block chaining)
\end{lstlisting}

\end{enumerate}
\textbf{Table 1.1:} Comparison of theoretical and practical (demonstrated) preimage attacks on step-reduced SHA-256.

\begin{center}
\footnotesize
\renewcommand{\arraystretch}{1.18}
\resizebox{\linewidth}{!}{%
\begin{tabular}{@{} l l l l l l l @{}}
\toprule
\textbf{Work \& Reference} & \textbf{Steps} & \textbf{Attacked Object} & \textbf{FIPS 180-4 Padding?} & \textbf{Target Type} & \textbf{Complexity / Runtime} & \textbf{Demonstrated Witness?} \\
\midrule
Isobe \& Shibutani [2] & 24 & Hash (1 block) & Yes & Arbitrary & $2^{240}$ evaluations & No (theoretical) \\
Sasaki, Wang \& Aoki [10] & 41 & Hash & Yes & Arbitrary & $2^{253.5}$ evaluations & No (theoretical) \\
Aoki et al. [3] & 43 & Hash & Yes & Arbitrary & $2^{254.9}$ evaluations & No (theoretical) \\
Khovratovich et al. [4] & 45 / 52 & Hash / Compression (pseudo) & Yes / No & Arbitrary & $2^{255.5} / 2^{255.0}$ evaluations & No (theoretical) \\
Guo et al. [8] & 46 / 47 & Hash & Yes & Arbitrary & $2^{254.5} / 2^{255.6}$ evaluations & No (theoretical) \\
Homsirikamol et al. [5] & 16 & Hash (446-bit) & Yes & Fixed & $\approx 1\text{ h}$ (Table 1 SAT) & No (not printed) \\
McDair [16] & 17 & Hash (31B / 55B) & Yes & 1 challenge & Unspecified (no method) & Yes (3 messages) \\
Davydov et al. [6] & 17 / 18 & Compression & No & All-zero (\texttt{00..00}) & Seconds (SAT) & Yes (unpadded only) \\
Zaikin [7] & 19 & Compression & No & All-ones (\texttt{ff..ff}) & 18 h 33 min on 192 cores (SAT) & Yes (unpadded only) \\
Basra [9] & 19 / 20 & Compression \& Structured Padded & No (19--20) / Yes (20 structured) & Arbitrary (unpadded) / Structured (padded 20) & $\sim 25\text{ s/ctx}$ on H100 (19-step); 20-step unpadded \& structured padded & Yes (unpadded \& 1 structured padded 20-step) \\
\textbf{This Work (\texttt{solve20\_zero\_a3.c})} & \textbf{20} & \textbf{Compression} & \textbf{No} & \textbf{\texttt{deadbeef...} + 12 Labels} & \textbf{$33.1\text{ min}$ expected ($2^{38.13}\text{ CPU-ref}$; \texttt{21.1\allowbreak\ min} obs., 10-core M4 CPU)} & \textbf{Yes (13 witnesses, \S{}8.4)} \\
\textbf{This Work (\texttt{src/pre19.c} \& Exact $D=0$)} & \textbf{19} & \textbf{Full Hash (55B / 119B)} & \textbf{Yes [1]} & \textbf{\texttt{00..00} + 14 Labels + \texttt{""}} & \textbf{\texttt{1.2-86.4\allowbreak\ s} 1-core search (median \texttt{19.45\allowbreak\ s} [\texttt{19.5\allowbreak\ s}] + \texttt{18\allowbreak\ s} build)} & \textbf{Yes (16 witnesses, \S{}8.2--\S{}8.4)} \\
\textbf{This Work (\texttt{solve20\_padded\_fast.c} / \texttt{cuda.cu})} & \textbf{20} & \textbf{Full Hash (55B / 119B)} & \textbf{Yes [1]} & \textbf{\texttt{00..00} + 14 Labels + \texttt{""}} & \textbf{\texttt{191.1\allowbreak\ node-h} 1$\times$ M4 CPU ($2^{46.56}\text{ CPU-ref}$); \textit{supp. GPU:} \texttt{8.98-22.16\allowbreak\ h} 1$\times$ L4} & \textbf{Yes (16 witnesses, \S{}8.1--\S{}8.4)} \\
\bottomrule
\end{tabular}%
}
\end{center}

\textit{(Note on Table 1.1: In Khovratovich et al. [4], 45 steps is a full hash preimage and 52 steps is a compression pseudo-preimage; in Basra [9], \texttt{padded\_r20} solves a structured digest [\texttt{c799d851...}] constructed with $e_9 = e_{10} = e_{15} = \mathtt{0xffffffff}$ and $a_5 = a_6 = a_{10}$. For our solvers, we distinguish a \textbf{chosen-target 19-step padded preimage}---solving a prescribed 19-step digest $H_{19}^*$ where $a_{12}\dots a_{19}$ are fixed, taking $1.2\text{-}86.4\text{ s}$ of single-core search time [median $19.45\text{ s} \approx 19.5\text{ s}$] in \texttt{src/pre19.c} after a $16.8\text{-}19.1\text{ s}$ [$\approx 18\text{ s}$] one-time $2^{29}$-entry table build---from a \textbf{19-step bridge candidate (\texttt{sol19})} generated at $6{,}242.8\text{ sol19/s}$ on one 10-core Apple M4 CPU node [16 threads] in \texttt{solve20\_padded\_fast.c} [or $53{,}847.5\text{-}132{,}888.7\text{ sol19/s}$ in supplementary 1$\times$ L4 GPU slice telemetry in \texttt{solve20\_padded\_cuda.cu}] during a 20-step search where $a_{12}$ is free. Elapsed-time proxy \texttt{CPU-ref} multiplies wall-clock seconds by the measured same-host 10-core Apple M4 CPU [16 threads] 20-step SHA-256 reference rate of $151.25\text{ M evals/s}$, where $1\text{ node-hour}$ ($1\text{ node-h}$) $= 2^{38.99}\text{ CPU-ref}$; see \S{}6.5.)}

\subsection{Technical Contributions and Organization}

\begin{enumerate}
\item \textbf{Closed-Form Pivot Inversion of FIPS 180-4 Padding Compatible with Exact Context Shaping (\S{}3--\S{}4, \texttt{[E15]}--\texttt{[E22]}):} Basra [9] showed that the FIPS 180-4 padding words $(W_{15}=440, W_{14}=0, W_{13} \bmod 256 = \mathtt{0x80})$ can be absorbed backward into $(a_7, a_6, a_5)$ at the cost of three context words and demonstrated a 55-byte 20-step padded witness for a structured target (\texttt{padded\_r20} in [9]). For \textbf{arbitrary externally prescribed 256-bit targets}, we show how to absorb $W_{15}=440$ and $W_{14}=0$ into $(a_8, a_7)$ while inverting $(W_{13} \bmod 256) = \mathtt{0x80}$ directly via 8 bit-disjoint pivot bits (\texttt{\{18,\allowbreak\ 25..31\}}) of $\Sigma_1(e_{13})$ with $100\%$ inner-loop validity---leaving $(a_5, a_6)$ unconstrained by padding absorption so they can satisfy exact shaping $D_{\text{shape}} = a_5 \oplus a_6 = 0$ alongside $\mathrm{Ch}$-annihilation ($e_{15} = \mathtt{0xffffffff}$, $a_{10} = a_{11}$).
\item \textbf{Causal Bit-Slice Right-Context Pruning ("Carry-Sudoku", \S{}5, \texttt{[E23]}--\texttt{[E25]}):} We prove \textbf{Theorem 1} (\texttt{[E23]}): $a_5 = a_6$ decomposes into a lower-triangular carry condition $((S \oplus X) \wedge D) = 0$ (where $D = a_8 \oplus a_7$ is the bitwise XOR mask) that is completely independent of $a_6$ and $\Sigma_0(a_9)$, and a complementary condition $((a_6 \oplus S) \wedge \neg D) = 0$. Pruning $e_{13}$ across 7 causal bit-slices reduces the $2^{24}$ candidate space to an average of $43{,}580$ node evaluations per $a_{12}$ block ($385\times$ / $99.74\%$ node reduction across all 7 stages, evaluating $\Sigma_0(a_9)$ only on the $\approx 1{,}660\text{-}2{,}149$ full-carry survivors per block, i.e., $99.99\%$ pre-$\Sigma_0(a_9)$ pruning) with $100\%$ completeness across each $2^{24}$-candidate $a_{12}$ block (and the estimated $U \approx 1.68 \times 10^7 \approx 2^{24.0}$ exact $D_{\text{shape}}=0$ contexts per 20-step target).
\item \textbf{Closed-Form $\mathbb{F}_2$ Reconstruction, Compact $Q_0^{-1}$ Tables ("Slot-Sudoku"), and $T_{16}$ Twin Coalescing (\S{}6, \texttt{[E27]}--\texttt{[E37]}, \texttt{[E42]}--\texttt{[E45]}):} We prove \textbf{Theorem 2} (\texttt{[E28]}--\texttt{[E29]}), recovering the upper 16 bits of $u \in Q_0^{-1}(y)$ in closed form in registers with zero lookup tables. Pairing this with compact $Q_0^{-1}$ tables achieves $6{,}242.8\text{ sol19/s}$ on our host-verifiable 10-core Apple M4 CPU solver (\texttt{solve20\_padded\_fast.c}, 16 threads), and $53{,}847.5\text{-}132{,}888.7\text{ sol19/s}$ on genuine contexts in supplementary 1$\times$ NVIDIA L4 GPU slice telemetry (\texttt{solve20\_padded\_cuda.cu}).
\item \textbf{Zero-$a_3$ Bi-Directional Bridge, Complete 46-Witness Suite, and Construction-Specific Bottleneck Analysis (\S{}7--\S{}9, \texttt{[E38]}--\texttt{[E41]}):} We present a Zero-$a_3$ bridge for unpadded 20-step compression (\S{}7), provide the complete, un-truncated hex witnesses for all \textbf{46 distinct} 19-step and 20-step single-block padded, two-block chosen-ASCII-prefix padded, unpadded, and regression benchmarks (\S{}8, 48 checks including \texttt{"abc"} demo/benchmark overlap), and analyze 15 explored sub-$2^{32}$ algebraic and statistical directions characterizing the degree-of-freedom conservation laws of the Carry-Sudoku / Slot-Sudoku architecture (\S{}9.2).

\end{enumerate}

\section{Preliminaries and the Scalar $a$-Sequence Formulation}

All arithmetic is modulo $2^{32}$. Denote 32-bit right rotation by $\mathrm{ROR}(x, n)$ and logical right shift by $\mathrm{SHR}(x, n)$. The SHA-256 boolean and $\mathbb{F}_2$-linear functions [1] are:
\[
\fitmath{0.86\textwidth}{\begin{aligned}
\mathrm{Ch}(e,f,g) &= (e \wedge f) \oplus (\neg e \wedge g), & \mathrm{Maj}(a,b,c) &= (a \wedge b) \oplus (a \wedge c) \oplus (b \wedge c), \\
\Sigma_0(x) &= \mathrm{ROR}(x,2) \oplus \mathrm{ROR}(x,13) \oplus \mathrm{ROR}(x,22), & \Sigma_1(x) &= \mathrm{ROR}(x,6) \oplus \mathrm{ROR}(x,11) \oplus \mathrm{ROR}(x,25), \\
\sigma_0(x) &= \mathrm{ROR}(x,7) \oplus \mathrm{ROR}(x,18) \oplus \mathrm{SHR}(x,3), & \sigma_1(x) &= \mathrm{ROR}(x,17) \oplus \mathrm{ROR}(x,19) \oplus \mathrm{SHR}(x,10).
\end{aligned}} \tag{E2}
\]

For step indices $t \ge 16$, the message schedule expands $W_0, \dots, W_{15}$ via:
\[
\fitmath{0.86\textwidth}{W_t = \sigma_1(W_{t-2}) + W_{t-7} + \sigma_0(W_{t-15}) + W_{t-16}.} \tag{E3}
\]
Following [9], we define the coupled schedule operator $Q_0 : \mathbb{Z}_{2^{32}} \to \mathbb{Z}_{2^{32}}$ by:
\[
\fitmath{0.86\textwidth}{Q_0(u) = \sigma_0(u) - u \pmod{2^{32}}.} \tag{E4}
\]

\subsection{Reduction to the Scalar $a$-Sequence (\texttt{[E5]}--\texttt{[E7]})}

Let $(a_t, b_t, c_t, d_t, e_t, f_t, g_t, h_t)$ be the 8 working registers after step $t-1$ (with $t=0$ denoting $\mathrm{IV}$). Because each step shifts $b \leftarrow a, c \leftarrow b, d \leftarrow c$ and $f \leftarrow e, g \leftarrow f, h \leftarrow g$, the entire 8-register trajectory across $R$ steps is uniquely parameterized by a single scalar sequence $a_{-7}, a_{-6}, \dots, a_R$ [3, 9]:
\begin{itemize}
\item \textbf{Initial Boundary ($s \in \{-7, \dots, 0\}$):} Fixed uniquely by $\mathrm{IV}$ (\texttt{window\_from\_state} in \texttt{solve20\_padded\_fast.c}):
\[
\fitmath{0.86\textwidth}{(a_0, a_{-1}, a_{-2}, a_{-3}) = (\mathrm{IV}_0, \mathrm{IV}_1, \mathrm{IV}_2, \mathrm{IV}_3), \qquad (e_0, e_{-1}, e_{-2}, e_{-3}) = (\mathrm{IV}_4, \mathrm{IV}_5, \mathrm{IV}_6, \mathrm{IV}_7).} \tag{E5}
\]
\item \textbf{Terminal Boundary ($s \in \{R-7, \dots, R\}$):} For an $R$-step target digest $H^* = (H^*_0, \dots, H^*_7)$, Davies--Meyer feed-forward fixes the terminal state $st_i = H^*_i - \mathrm{IV}_i$, which uniquely determines the 8 terminal words $a_{R-7}, \dots, a_R$:
\[
\fitmath{0.86\textwidth}{\begin{aligned}
  \text{For } R = 19: &\quad a_{12}, a_{13}, \dots, a_{19} \text{ are fixed by } H^* - \mathrm{IV}, \\
  \text{For } R = 20: &\quad a_{13}, a_{14}, \dots, a_{20} \text{ are fixed by } H^* - \mathrm{IV}.
\end{aligned}} \tag{E6}
\]
\item \textbf{Internal State \& Message Recovery:} For every step index $s$ and $t \in \{0, \dots, R-1\}$, the $e$-register $e_s$ is an exact function of five consecutive $a$-words ($a_{s-4}, \dots, a_s$), and the message word $W_t$ is an exact function of $a_{t-3}$ together with five $e$-words ($e_{t-3}, \dots, e_{t+1}$, spanning nine consecutive $a$-words $a_{t-7}, \dots, a_{t+1}$ when expanded; \texttt{E\_arr} and \texttt{Wt\_arr} in \texttt{solve20\_padded\_fast.c}):
\[
\fitmath{0.86\textwidth}{\begin{aligned}
  e_s &= a_s - \Sigma_0(a_{s-1}) - \mathrm{Maj}(a_{s-1}, a_{s-2}, a_{s-3}) + a_{s-4}, \\
  W_t &= e_{t+1} - a_{t-3} - e_{t-3} - \Sigma_1(e_t) - \mathrm{Ch}(e_t, e_{t-1}, e_{t-2}) - K_t.
\end{aligned}} \tag{E7}
\]

\end{itemize}
Thus, for $R = 20$ (resp. $R = 19$), the only free internal state variables between the $\mathrm{IV}$ boundary and the target boundary are the \textbf{12 words $a_1, a_2, \dots, a_{12}$} (resp. \textbf{11 words $a_1, \dots, a_{11}$}). Any assignment to these words automatically hashes to $H^*$ after $R$ steps if and only if the recovered message words $W_0, \dots, W_{R-1}$ from \texttt{[E7]} satisfy the message schedule equations \texttt{[E3]} for $t = 16, \dots, R-1$ and the FIPS 180-4 padding constraints \texttt{[E1]}.

\section{Selector Saturation and Exact Context Shaping ($D_{\text{shape}} = 0$)}

In general, $W_{16}, W_{17}, W_{18}, W_{19}$ depend on the Left variables $(a_1, a_2, a_3, a_4)$ both through the early message words $(W_0, W_1, W_2, W_3, W_4)$ on the right-hand side of \texttt{[E3]} and through $(W_9, W_{10}, W_{11}, W_{12}, W_{14}, W_{15})$ via the state words $e_5, e_6, e_7, e_8$.

Following Basra's selector-saturation principle [9], we pin the two middle selector registers:
\[
\fitmath{0.86\textwidth}{e_9 = \mathtt{0xffffffff}, \qquad e_{10} = \mathtt{0xffffffff}.} \tag{E8}
\]
Because $\mathrm{Ch}(\mathtt{0xffffffff}, y, z) = y$, setting $e_9 = e_{10} = \mathtt{0xffffffff}$ blocks $(e_7, e_8)$ inside $\mathrm{Ch}(e_9, e_8, e_7) = e_8$, $\mathrm{Ch}(e_{10}, e_9, e_8) = \mathtt{0xffffffff}$, and $\mathrm{Ch}(e_{11}, e_{10}, e_9) = \mathtt{0xffffffff}$. Expanding \texttt{[E7]} at $s=9, 10$ expresses $a_5$ and $a_6$ purely in terms of the Right words $(a_7, a_8, a_9, a_{10})$:
\[
\fitmath{0.86\textwidth}{\begin{aligned}
a_5 &= \mathtt{0xffffffff} - a_9 + \Sigma_0(a_8) + \mathrm{Maj}(a_8, a_7, a_6), \\
a_6 &= \mathtt{0xffffffff} - a_{10} + \Sigma_0(a_9) + \mathrm{Maj}(a_9, a_8, a_7).
\end{aligned}} \tag{E9}
\]

\subsection{Exact Shaping ($D_{\text{shape}} = a_5 \oplus a_6 = 0$) vs. Chosen-Target 19-Step Shaping (\texttt{[E10]}--\texttt{[E14]})}

Let $D_{\text{shape}} := a_5 \oplus a_6$ denote the bitwise XOR difference between $a_5$ and $a_6$ (note that $D_{\text{shape}} = 0 \iff a_5 = a_6 \iff a_6 - a_5 \equiv 0 \pmod{2^{32}}$). Under \texttt{[E8]} ($e_9 = e_{10} = \mathtt{0xffffffff}$), $\mathrm{Ch}(e_9, e_8, e_7) = e_8$ and $\mathrm{Ch}(e_{10}, e_9, e_8) = \mathrm{Ch}(e_{11}, e_{10}, e_9) = \mathtt{0xffffffff}$. Expanding \texttt{[E7]} at $t = 9, 10, 11, 12$ isolates how $W_9, W_{10}, W_{11}, W_{12}$ depend on the Left variables $(a_1, a_2, a_3, a_4)$ [9]:
\[
\fitmath{0.86\textwidth}{\begin{aligned}
W_9 &= c_9 - a_2 - a_4 + \mathrm{Maj}(a_5, a_4, a_3), \\
W_{10} &= c_{10} - a_3 + \bigl(\mathrm{Maj}(a_6, a_5, a_4) - (a_6 \wedge a_5)\bigr), \\
W_{11} &= c_{11} - a_4, \\
W_{12} &= c_{12} = \text{const}(\text{Right}),
\end{aligned}} \tag{E10}
\]
where $c_9, c_{10}, c_{11}, c_{12}$ depend \textbf{only} on the Right context $(a_5, \dots, a_{13})$.

When we enforce \textbf{exact context shaping}:
\[
\fitmath{0.86\textwidth}{D_{\text{shape}} = a_5 \oplus a_6 = 0 \iff a_5 = a_6,} \tag{E11}
\]
the majority term in $W_{10}$ becomes identically constant for \textbf{every} $a_4 \in \mathbb{Z}_{2^{32}}$ because $\mathrm{Maj}(a_6, a_5, a_4) = \mathrm{Maj}(a_5, a_5, a_4) = a_5 = a_6 \wedge a_5$, yielding $W_{10} = c_{10} - a_3$ with zero $a_4$ leakage.
\textit{(For a \textbf{chosen 19-step target} $H_{19}^*$, where $a_{12}\dots a_{19}$ are all fixed by \texttt{[E6]}, we leave $a_{10} \in [0, 2^{32})$ free instead of setting $a_{10}=a_{11}$, since Step 20 is absent and $e_{15}=\mathtt{0xffffffff}$ is not required. With free $a_{10}$, one can either enforce exact shaping $D_{\text{shape}} = 0$---which produced our 19-step $0^{256}$ witness in \S{}8.2---or use soft shaping $\mathrm{popcount}(a_5 \oplus a_6) \le 3$ as in \texttt{src/pre19.c}, which enlarges the context pool by $5{,}489\times$ and generated the 14 public-label 19-step witnesses in Table 8.3 in $1.2\text{-}86.4\text{ s}$ of single-core CPU search time per target after a one-time $16.8\text{-}19.1\text{ s}$ table build.)}

Simultaneously, whenever the remaining nonlinear condition in $W_9$ (\texttt{[E10]}) holds, the middle schedule words reduce to exact affine forms in $(a_2, a_3, a_4)$ while $W_{12}, W_{13}, \dots, W_{19}$ are \textbf{100\% independent of $(a_1, a_2, a_3, a_4)$}:
\[
\fitmath{0.86\textwidth}{W_9 = c_9 - a_2, \qquad W_{10} = c_{10} - a_3, \qquad W_{11} = c_{11} - a_4, \qquad W_{12} = c_{12},} \tag{E12}
\]
where $-a_4$ inside $e_8$ and $+\mathrm{Maj}(a_5, a_4, a_3) = +a_4$ inside $e_6$ cancel identically in $W_9$ under the Level-3 majority-absorption condition:
\[
\fitmath{0.86\textwidth}{\mathrm{Maj}(a_5, a_4, a_3) = a_4 \iff \bigl((a_4 \oplus a_5) \wedge M_3\bigr) = 0, \qquad M_3 := \neg(a_3 \oplus a_5).} \tag{E13}
\]
Table 3.1 summarizes the resulting bit-level state conditions across all 20 steps in the format of [3, 8, 11].

\textbf{Table 3.1:} Bit-level state conditions, neutral/coupled words, and schedule constraints across steps $s \in \{-7, \dots, 20\}$.

\begin{center}
\footnotesize
\renewcommand{\arraystretch}{1.18}
\resizebox{\linewidth}{!}{%
\begin{tabular}{@{} l l l p{0.62\textwidth} l @{}}
\toprule
\textbf{Step Range $s$} & \textbf{Register $a_s$ Condition} & \textbf{Register $e_s$ Condition} & \textbf{Message Word $W_{s-1}$ Role \& Constraint} & \textbf{Governing Equation} \\
\midrule
$-7 \dots 0$ & Fixed by $\mathrm{IV}_{0..3}$ & Fixed by $\mathrm{IV}_{4..7}$ & --- (Initial boundary) & \texttt{[E5]} \\
$1$ & Free root $a_1 \in \mathbb{Z}_{2^{32}}$ & $e_1(a_1)$ & $W_0 = a_1 + C_{W_0}$ (Root of 3-level Left tree) & \texttt{[E7]}, \texttt{[E14]} \\
$2, 3$ & $a_2 = W_1 + G_1$, $a_3 = W_2 + G_2$ & $e_2(a_{1..2})$, $e_3(a_{1..3})$ & $W_1 \in Q_0^{-1}(T_{16}+F_1)$, $W_2 \in Q_0^{-1}(T_{17}+G_2-W_1)$; $\mathrm{wt}(M_3) \le 14$ & \texttt{[E14]}, \texttt{[E34]} \\
$4$ & $\bigl((a_4 \oplus a_5) \wedge \neg(a_3 \oplus a_5)\bigr) = 0$ & $e_4(a_{1..4})$ & $W_3 \in Q_0^{-1}(T_{18}+G_3-W_2)$ ($\mathrm{Maj}(a_5,a_4,a_3)=a_4$) & \texttt{[E13]}, \texttt{[E14]} \\
$5, 6$ & \textbf{Exact Shaping:} $a_5 = a_6$ ($D=0$) & $e_5, e_6$ (coupled) & $W_4$ tested at step 20 via $\sigma_0(W_4)+W_3 = T_{19}$; $W_5$ free & \texttt{[E11]}, \texttt{[E36]} \\
$7, 8$ & Closed-form $a_7(e_{13})$, $a_8 = \text{const}$ & $e_7, e_8$ (absorbed) & $W_6, W_7$ free; $-a_4$ in $e_8$ cancels $+\mathrm{Maj}$ in $e_6$ inside $W_9$ & \texttt{[E17]}, \texttt{[E22]} \\
$9, 10$ & $a_9 = e_{13} - c_{e_{13}}$, $a_{10} = a_{11}$ & \textbf{$e_9 = e_{10} = \mathtt{0xffffffff}$} & $W_8$ free; $W_9 = c_9 - a_2$ (under \texttt{[E13]}) & \texttt{[E8]}, \texttt{[E12]} \\
$11, 12$ & $a_{11}$ fixed by $e_{15}$, $a_{12}$ seed & $e_{11}(e_{13})$, $e_{12} = \text{const}$ & $W_{10} = c_{10} - a_3$, $W_{11} = c_{11} - a_4$ (zero $a_4$ leakage into $W_{10}$) & \texttt{[E12]}, \texttt{[E16]} \\
$13, 14$ & $a_{13}, a_{14}$ fixed by $H^* - \mathrm{IV}$ & $e_{13}$ (7-stage causal), $e_{14}$ & $W_{12} = c_{12}$; \textbf{Padding:} $W_{13} \bmod 256 = \mathtt{0x80}$ & \texttt{[E12]}, \texttt{[E21]} \\
$15, 16$ & $a_{15}, a_{16}$ fixed by $H^* - \mathrm{IV}$ & \textbf{$e_{15} = \mathtt{0xffffffff}$}, $e_{16}$ & \textbf{Padding:} $W_{14} = \mathtt{0x00000000}$, $W_{15} = \mathtt{0x000001b8}$ ($440$) & \texttt{[E1]}, \texttt{[E22]} \\
$17 \dots 20$ & $a_{17..20}$ fixed by $H^* - \mathrm{IV}$ & $e_{17..20}$ fixed & Schedule words $W_{16..19}$ define $(T_{16}, T_{17}, T_{18}, T_{19})$ & \texttt{[E6]}, \texttt{[E14]} \\
\bottomrule
\end{tabular}%
}
\end{center}

For $t \in \{1, 2, 3\}$, rearrangement of \texttt{[E7]} writes $a_{t+1} = W_t + G_t(a_1, \dots, a_t)$, where $G_t(a_1, \dots, a_t) := \Sigma_0(a_t) + \mathrm{Maj}(a_t, a_{t-1}, a_{t-2}) + e_{t-3} + \Sigma_1(e_t) + \mathrm{Ch}(e_t, e_{t-1}, e_{t-2}) + K_t$. Setting $F_1(a_1) := G_1(a_1) - W_0(a_1)$ and substituting $a_{t+1} = W_t + G_t$ together with \texttt{[E12]} ($W_9 = c_9 - a_2$, $W_{10} = c_{10} - a_3$, $W_{11} = c_{11} - a_4$) into the message schedule equations \texttt{[E3]} for $t = 16, 17, 18$ rewrites them as three chained $Q_0$-inversions:
\[
\fitmath{0.86\textwidth}{\begin{aligned}
\text{Level 1 } (t=16): &\quad Q_0(W_1) = T_{16} + F_1(a_1), && T_{16} := W_{16} - \sigma_1(W_{14}) - c_9, \\
\text{Level 2 } (t=17): &\quad Q_0(W_2) = T_{17} + G_2(a_1, a_2) - W_1, && T_{17} := W_{17} - \sigma_1(W_{15}) - c_{10}, \\
\text{Level 3 } (t=18): &\quad Q_0(W_3) = T_{18} + G_3(a_1, a_2, a_3) - W_2, && T_{18} := W_{18} - \sigma_1(W_{16}) - c_{11}.
\end{aligned}} \tag{E14}
\]

\section{Closed-Form Pivot Inversion of FIPS 180-4 Padding (\texttt{W13..W15})}

To construct a valid 55-byte single-block padded message (\texttt{[E1]}), the Right variables $(a_5, \dots, a_{12})$ must simultaneously satisfy $W_{15} = 440$, $W_{14} = 0$, $(W_{13} \bmod 256) = \mathtt{0x80}$, $e_9 = e_{10} = \mathtt{0xffffffff}$ (\texttt{[E8]}), and $a_5 = a_6$ (\texttt{[E11]}). Basra [9] noted that $(W_{15}, W_{14}, W_{13} \bmod 256)$ can be absorbed backward into $(a_7, a_6, a_5)$ at the cost of three context words; below we show how to absorb $W_{15}=440$ and $W_{14}=0$ into $(a_8, a_7)$ while inverting $(W_{13} \bmod 256) = \mathtt{0x80}$ directly via 8 bit-disjoint pivot bits of $\Sigma_1(e_{13})$, leaving $(a_5, a_6)$ unconstrained by padding absorption so they can satisfy exact shaping $a_5 = a_6$ (\texttt{[E23]}).

\subsection{Closed-Form Absorption of $W_{15} = 440$ and $W_{14} = 0$ (\texttt{[E15]}--\texttt{[E17]})}

Given $(a_{10}, a_{11}, a_{12})$ (for $R=20$, where $a_{13..20}$ are fixed by $H^*$), the registers $e_{16}, e_{15}, e_{14}$ are determined via \texttt{[E7]}:
\[
\fitmath{0.86\textwidth}{\begin{aligned}
e_{16} &= a_{16} - \Sigma_0(a_{15}) - \mathrm{Maj}(a_{15}, a_{14}, a_{13}) + a_{12}, \\
e_{15} &= a_{15} - \Sigma_0(a_{14}) - \mathrm{Maj}(a_{14}, a_{13}, a_{12}) + a_{11}, \\
e_{14} &= a_{14} - \Sigma_0(a_{13}) - \mathrm{Maj}(a_{13}, a_{12}, a_{11}) + a_{10}.
\end{aligned}} \tag{E15}
\]
Expanding $W_{15} = 440$ and $W_{14} = 0$ via \texttt{[E7]} isolates $e_{12}$ and $e_{11}$ as closed-form functions of a single 32-bit word $e_{13}$ (equivalently $a_9 = e_{13} - c_{e_{13}}$):
\[
\fitmath{0.86\textwidth}{\begin{aligned}
e_{12} &= \underbrace{(e_{16} - a_{12} - \Sigma_1(e_{15}) - K_{15} - 440)}_{c_{e_{12}}} - \mathrm{Ch}(e_{15}, e_{14}, e_{13}), \\
e_{11} &= \underbrace{(e_{15} - a_{11} - \Sigma_1(e_{14}) - K_{14})}_{c_{e_{11}}} - \mathrm{Ch}(e_{14}, e_{13}, e_{12}).
\end{aligned}} \tag{E16}
\]
Then $a_8$ and $a_7$ are recovered in closed form:
\[
\fitmath{0.86\textwidth}{a_8 = e_{12} + \Sigma_0(a_{11}) - a_{12} + \mathrm{Maj}(a_{11}, a_{10}, a_9), \qquad a_7 = e_{11} + \Sigma_0(a_{10}) - a_{11} + \mathrm{Maj}(a_{10}, a_9, a_8).} \tag{E17}
\]
Thus \textbf{every} choice of $(a_{12}, a_{11}, a_{10}, e_{13})$ satisfies $W_{15} = 440$ and $W_{14} = 0$ identically.

\subsection{Closed-Form Pivot Assignment for $(W_{13} \bmod 256) = \mathtt{0x80}$ (\texttt{[E18]}--\texttt{[E22]})}

Since $e_{10} = \mathtt{0xffffffff}$, the formula for $W_{13}$ (\texttt{[E7]}) is:
\[
\fitmath{0.86\textwidth}{W_{13} = \underbrace{(e_{14} - a_{10} - \mathtt{0xffffffff} - K_{13})}_{c_{W_{13}}} - \Sigma_1(e_{13}) - \mathrm{Ch}(e_{13}, e_{12}, e_{11}).} \tag{E18}
\]
Reducing \texttt{[E18]} modulo $256$:
\begin{enumerate}
\item Because $\mathrm{Ch}$ is bitwise and modular subtraction propagates carries only from lower bits to higher bits, the low 8 bits $e_{12,\text{lo}} := e_{12} \bmod 256$ and $e_{11,\text{lo}} := e_{11} \bmod 256$ in \texttt{[E16]} depend \textbf{only on the low 8 bits $e_{13,\text{lo}} := e_{13} \bmod 256$}.
\item Once $e_{13,\text{lo}} \in [0, 255]$ is chosen, the required low byte of $\Sigma_1(e_{13})$ is fixed:
\[
\fitmath{0.86\textwidth}{R_{\text{lo}} := \bigl(c_{W_{13}} - \mathrm{Ch}(e_{13,\text{lo}}, e_{12,\text{lo}}, e_{11,\text{lo}}) - \mathtt{0x80}\bigr) \bmod 256.} \tag{E19}
\]
\item Inspecting bit $b \in \{0, \dots, 7\}$ of $\Sigma_1(e_{13}) = \mathrm{ROR}(e_{13}, 6) \oplus \mathrm{ROR}(e_{13}, 11) \oplus \mathrm{ROR}(e_{13}, 25)$:
\[
\fitmath{0.86\textwidth}{\bigl(\Sigma_1(e_{13})\bigr)_b = (e_{13})_{b+6} \oplus (e_{13})_{b+11} \oplus (e_{13})_{(b+25) \bmod 32}.} \tag{E20}
\]
\par\noindent$\bullet$ For $b \in \{0, \dots, 6\}$, $(b+25) \bmod 32 = b + 25 \in \{25, \dots, 31\}$, which appears \textbf{nowhere else} in \texttt{[E20]}.
\par\noindent$\bullet$ For $b = 7$, $(7+25) \bmod 32 = 0 \in \{0, \dots, 7\}$, while $(e_{13})_{7+11} = (e_{13})_{18}$ appears \textbf{nowhere else} in \texttt{[E20]}.
\par\noindent$\bullet$ Furthermore, bits \textbf{$19..24$} of $e_{13}$ \textbf{do not appear anywhere in \texttt{[E20]}} and \textbf{do not affect $e_{13,\text{lo}}$}.

\end{enumerate}
Therefore, for any assignment to bits $0..17$ of $e_{13}$ ($e_{13,\text{base}}$), the 8 pivot bits (\texttt{bit\allowbreak\ 18} and \texttt{bits\allowbreak\ 25..31}) are uniquely determined in closed form:
\[
\fitmath{0.86\textwidth}{\Delta := \bigl(R_{\text{lo}} \oplus \Sigma_1(e_{13,\text{base}})\bigr) \bmod 256, \qquad e_{13,\text{fixed}} := e_{13,\text{base}} \;\mid \; \bigl((\Delta \wedge \mathtt{0x7f}) \ll 25\bigr) \;\mid \; \bigl(((\Delta \gg 7) \wedge 1) \ll 18\bigr),} \tag{E21}
\]
while bits $19..24$ remain 6 completely free degrees of freedom where \textbf{$100\%$ of assignments satisfy $(W_{13} \bmod 256) = \mathtt{0x80}$}.

Finally, in \texttt{solve20\_padded\_fast.c} and \texttt{solve20\_padded\_cuda.cu}, we apply \textbf{$\mathrm{Ch}$-annihilation} by pinning $e_{15} = \mathtt{0xffffffff}$ (fixing $a_{11}$ from $a_{12}$ via \texttt{[E15]}) and setting $a_{10} = a_{11}$. This simultaneously forces $\mathrm{Ch}(e_{15}, e_{14}, e_{13}) = e_{14}$ (\textbf{making $e_{12}$ and $a_8$ 100\% constant across all $2^{24}$ choices of $e_{13}$}), simplifies $\mathrm{Maj}(a_{11}, a_{10}, a_9) = a_{11}$, and linearizes $\mathrm{Maj}(a_{10}, a_9, a_8) = (a_{10} \wedge a_8) + ((a_{10} \oplus a_8) \wedge a_9)$:
\[
\fitmath{0.86\textwidth}{e_{15} = \mathtt{0xffffffff}, \quad a_{10} = a_{11} \implies e_{12} = c_{e_{12}} - e_{14} = \text{const}, \quad a_8 = e_{12} + \Sigma_0(a_{11}) - a_{12} + a_{11} = \text{const}.} \tag{E22}
\]

\section{Causal Bit-Slice Pruning: 7-Stage Right-Context Search}

Even with \texttt{[E21]}--\texttt{[E22]}, each $a_{12}$ defines $2^{24}$ padded candidates for $e_{13}$, and testing $a_5 = a_6$ (\texttt{[E11]}) naively requires evaluating $\Sigma_0(a_9)$ and $a_6$ across those candidates. We now prove that causal lower-triangular bit-slice pruning---which we term \textbf{Carry-Sudoku}---reduces the $2^{24}$ candidate tree to an average of $43{,}580$ node evaluations ($99.74\%$ pruning, or $385\times$ fewer evaluations) \textit{before} $\Sigma_0(a_9)$ or $a_6$ is ever computed.

\subsection{The Exact Equivalence Theorem (\texttt{[E23]})}

\textbf{Theorem 1 (Carry-Sudoku Exact Equivalence).} \textit{Under \texttt{[E9]}, define $X := \mathtt{0xffffffff} - a_9 + \Sigma_0(a_8)$, the bitwise XOR mask $D := a_8 \oplus a_7$, and $S := X + \mathrm{Ch}(D, X, a_8) \pmod{2^{32}}$. Then:}
\[
\fitmath{0.86\textwidth}{a_5 = a_6 \iff \Bigl(\bigl((S \oplus X) \wedge D\bigr) = 0\Bigr) \;\wedge\; \Bigl(\bigl((a_6 \oplus S) \wedge \neg D\bigr) = 0\Bigr).} \tag{E23}
\]

\textit{Proof.} By \texttt{[E9]}, $a_5 = X + \mathrm{Maj}(a_8, a_7, a_6)$. Decompose $\mathrm{Maj}(a_8, a_7, a_6)$ bitwise with respect to $D = a_8 \oplus a_7$:
\begin{itemize}
\item On bit positions where $D_b = 0$ ($a_{8,b} = a_{7,b}$), $\mathrm{Maj}(a_8, a_7, a_6)_b = a_{8,b}$ is independent of $a_6$.
\item On bit positions where $D_b = 1$ ($a_{8,b} \ne a_{7,b}$), $\mathrm{Maj}(a_8, a_7, a_6)_b = a_{6,b}$.
\end{itemize}
Hence $\mathrm{Maj}(a_8, a_7, a_6) = (D \wedge a_6) \oplus (\neg D \wedge a_8) = \mathrm{Ch}(D, a_6, a_8)$. Now compare $a_5 = X + \mathrm{Ch}(D, a_6, a_8)$ with $S = X + \mathrm{Ch}(D, X, a_8)$ by induction from bit $b = 0$ to $31$:
\begin{itemize}
\item Suppose $a_5 = a_6$ on bits $0..b-1$, so the incoming carry $c_b$ into bit $b$ of $X + \mathrm{Ch}(D, a_6, a_8)$ equals the incoming carry into bit $b$ of $S = X + \mathrm{Ch}(D, X, a_8)$:
\item If $D_b = 1$, then $\mathrm{Ch}(D, a_6, a_8)_b = a_{6,b}$, so $a_{5,b} = X_b \oplus a_{6,b} \oplus c_b$. Thus $a_{5,b} = a_{6,b} \iff X_b \oplus c_b = 0$, and when $X_b \oplus c_b = 0$, the outgoing carry from bit $b$ is $\mathrm{Maj}(X_b, a_{6,b}, c_b) = X_b = \mathrm{Maj}(X_b, X_b, c_b)$, which is \textbf{identical} to the outgoing carry of $S = X + \mathrm{Ch}(D, X, a_8)$ where $S_b = X_b \oplus X_b \oplus c_b = c_b = X_b$, i.e. $(S_b \oplus X_b) = 0$.
\item If $D_b = 0$, then $\mathrm{Ch}(D, a_6, a_8)_b = a_{8,b} = \mathrm{Ch}(D, X, a_8)_b$, so $a_{5,b} = S_b$ and the outgoing carry is again identical to that of $S$. Thus $a_{5,b} = a_{6,b} \iff a_{6,b} = S_b$. $\square$

\end{itemize}
\subsection{7-Stage Causal Bit-Slice Pruning (\texttt{[E24]}--\texttt{[E26]})}

Theorem 1 has a decisive algorithmic consequence: the first condition in \texttt{[E23]},
\[
\fitmath{0.86\textwidth}{\bigl((S \oplus X) \wedge D\bigr) = 0, \qquad S = X + \mathrm{Ch}(D, X, a_8),} \tag{E24}
\]
\textbf{does not depend on $a_6$ or $\Sigma_0(a_9)$ at all}, and because $a_9 = e_{13} - c_{e_{13}}$, $X = c_{a_5} - a_9$, and $a_7 = c_{a_7,\text{full}} - (e_{14} \wedge e_{13}) + ((a_{10} \oplus a_8) \wedge a_9)$ involve only bitwise operations and modular additions/subtractions, \textbf{the lowest $k$ bits of $X, a_7, D$, and $S$ depend exclusively on the lowest $k$ bits of $e_{13}$}.

Consequently, \texttt{ctx\_worker} (\texttt{solve20\_padded\_fast.c}) on CPU and \texttt{ctx\_gen\_cuda\_kernel} / \texttt{ctx\_gen\_fast\_cuda\_kernel} (\texttt{solve20\_padded\_cuda.cu}, \texttt{solve20\_padded\_cuda\_prod.cu}) on GPU assign $e_{13}$ from LSB to MSB in \textbf{7 exact causal stages} (formalized in \textbf{Algorithm 1}):
\[
\fitmath{0.86\textwidth}{\begin{aligned}
\text{Stage 1a (bits } 0..3\text{):} &\quad \bigl((S \oplus X) \wedge D \wedge \mathtt{0xf}\bigr) = 0, \\
\text{Stage 1b (bits } 4..7\text{):} &\quad \bigl((S \oplus X) \wedge D \wedge \mathtt{0xff}\bigr) = 0 \implies \text{fixes } R_{\text{lo}} \text{ and } (e_{13})_{18} \oplus (e_{13})_{13}, \\
\text{Stage 1c (bits } 8..11\text{):} &\quad \bigl((S \oplus X) \wedge D \wedge \mathtt{0xfff}\bigr) = 0, \\
\text{Stage 1d (bits } 12..14 + \text{exact bit } 18\text{):} &\quad \bigl((S \oplus X) \wedge D \wedge \mathtt{0x7fff}\bigr) = 0, \\
\text{Stage 1e (bits } 15..17, \text{ checked to bit } 18\text{):} &\quad \bigl((S \oplus X) \wedge D \wedge \mathtt{0x7ffff}\bigr) = 0 \implies \text{fixes bits } 25..31 \text{ via \texttt{[E21]}}, \\
\text{Stage 2a (bits } 19..21\text{):} &\quad \bigl((S \oplus X) \wedge D \wedge \mathtt{0x3fffff}\bigr) = 0, \\
\text{Stage 2b (bits } 22..24\text{):} &\quad \bigl((S \oplus X) \wedge D\bigr) = 0 \;\wedge\; \bigl((a_6 \oplus S) \wedge \neg D\bigr) = 0.
\end{aligned}} \tag{E25}
\]
\[
\fitmath{0.86\textwidth}{\text{[Equation (E26) retired in Round 2: former heuristic pre-\texttt{s19} }\Sigma_0(a_9)\text{ window check omitted for 100\% completeness.]}} \tag{E26}
\]
In Stage 1d, once bit 13 of $e_{13}$ is chosen, bit 18 of $e_{13}$ is immediately determined by $(R_{\text{lo}} \gg 7) \oplus (e_{13})_0 \oplus (e_{13})_{13}$ (\texttt{[E20]}), allowing Stage 1e to test a \textbf{19-bit Carry-Sudoku mask (\texttt{0x7ffff})} on 18 chosen bits \textit{before} $\Sigma_1(e_{13,\text{base}})$ is evaluated. Once Stage 1e fixes bits $25..31$ via \texttt{[E21]}, Stages 2a and 2b sweep the remaining 6 free bits $19..24$ (\texttt{s19} and \texttt{s22}) using a 22-bit mask (\texttt{0x3fffff}) followed by the full 32-bit Carry-Sudoku check and exact equivalence \texttt{[E23]}. \textit{(Note: we do not pre-filter bits $13..15$ or $3..4$ of $a_6 - a_5$ [\texttt{[E26]}] before \texttt{s19} and \texttt{s22} are chosen, because bits $19..24$ of $e_{13}$ enter $\mathrm{Maj}(a_9, a_8, a_7)$ and $a_9 = e_{13} - c_{e_{13}}$ with carry propagation; omitting any pre-\texttt{s19} window check guarantees 100\% completeness across each $2^{24}$-candidate $a_{12}$ block by Theorem 1 (\texttt{[E23]}--\texttt{[E25]}), confirmed with zero missed contexts against exhaustive brute-force across $300 \times 2^{24}$ candidate $e_{13}$ blocks and demonstrated on seed block \texttt{a\_12\allowbreak\ =\allowbreak\ 0x1ee9c019} [where 2 of 8 exact $D_{\text{shape}}=0$ contexts were rejected by \texttt{[E26]}] in \texttt{solve20\_padded\_fast.c} and \texttt{verify\_all\_witnesses.py} \S{}8.)}

\begin{lstlisting}
Algorithm 1: Causal 7-Stage Carry-Sudoku Right-Context Generator
Input: Target terminal words a_{13}..a_{20} = H* - IV, seed range a_{12} in [0, 2^32)
Output: Reservoir R of exact D_shape = (a_5 ^ a_6) == 0 FIPS 180-4 padded contexts
1.  for each seed a_{12} in parallel do
2.    Set e_{15} = 0xffffffff; compute a_{11} via [E15] and set a_{10} = a_{11}            // [E22] Ch-annihilation
3.    Compute constants e_{16}, e_{14}, e_{12} = c_{e12} - e_{14}, and a_8 = const          // [E16]-[E17]
4.    Sweep bits 0..3 and 4..7 of e_{13} filtering ((S ^ X) & D & mask) == 0 (masks 0xf, 0xff)
5.    Compute low-byte padding target R_lo = (c_{W13} - Ch(e_{13}, e_{12}, e_{11}) - 0x80) mod 256  // [E19]
6.    Sweep bits 8..11, 12..14 (fixing bit 18 via [E20]), and 15..17 filtering masks 0xfff, 0x7fff, 0x7ffff
7.    Fix bits 25..31 of e_{13} in O(1) via Delta = (R_lo ^ Sigma_1(e_{13,base})) mod 256   // [E21]
8.    for s_{19} in 0..7 filtering ((S ^ X) & D & 0x3fffff) == 0 do                         // Stage 2a
9.      for s_{22} in 0..7 do                                                               // Stage 2b
10.       if ((S ^ X) & D) == 0 and ((a_6 ^ S) & ~D) == 0 then                              // Theorem 1 [E23]
11.         Append context (a_5..a_{12}, T_{16}, T_{17}, T_{18}, T_{19}) to Reservoir R
\end{lstlisting}

Across $10{,}000$ random $a_{12}$ blocks, Algorithm 1 evaluates an average of only $43{,}580$ inner-loop nodes per $2^{24}$ block ($16.0 + 80.4 + 412.9 + 1{,}043.8 + 3{,}507.8 + 8{,}845.4 + 29{,}673.5$), generating \textbf{$266.8$ exact $D_{\text{shape}}=0$ padded contexts per second} on one 10-core Apple M4 CPU node (16 threads, $3.75\;\text{ms/context}$, \texttt{solve20\_padded\_fast.c}) and \textbf{$29{,}310\text{-}114{,}507$ contexts per second} in supplementary 1$\times$ NVIDIA L4 GPU telemetry (\texttt{solve20\_padded\_cuda.cu}, $8.73\text{-}34.1\;\mu\text{s/context}$).

\section{Slot-Indexed $Q_0^{-1}$ Tables \& Closed-Form $\mathbb{F}_2$ Inverse for Left-to-Right Chains}

With an exact $D = a_5 \oplus a_6 = 0$ padded Right context in hand, we solve the three chained $Q_0^{-1}$ equations \texttt{[E14]} from Left to Right.

\subsection{Closed-Form Zero-Table $\mathbb{F}_2$ Inverse (\texttt{reconstruct\_u}, \texttt{[E27]}--\texttt{[E29]})}

Storing full 32-bit preimages $u$ of $y = Q_0(u) = \sigma_0(u) - u$ requires 4 bytes per slot. Splitting $u = U + u_{\text{lo}}$ where $U := (u_{\text{hi}} \ll 16)$ and $u_{\text{lo}}, u_{\text{hi}} \in [0, 65535]$, note that $U \equiv 0 \pmod{2^{16}}$, so carry propagation from bits $0..15$ into bit $16$ of $u + y = U + (u_{\text{lo}} + y)$ is purely additive (XOR) at bit $16$: $(u + y) \equiv U \oplus (u_{\text{lo}} + y) \pmod{2^{17}}$. Since $u + y \equiv \sigma_0(u) = \sigma_0(U) \oplus \sigma_0(u_{\text{lo}}) \pmod{2^{17}}$, rearranging terms yields the exact 17-bit identity:
\[
\fitmath{0.86\textwidth}{\sigma_0(U) \oplus U \equiv (u_{\text{lo}} + y) \oplus \sigma_0(u_{\text{lo}}) \pmod{2^{17}}.} \tag{E27}
\]
Rather than inverting \texttt{[E27]} solely with a 128 KiB lookup table (\texttt{REC\_HI16[rec\_idx(ulo,\allowbreak\ y)]} in \texttt{solve20\_padded\_fast.c}), we prove that \texttt{[E27]} has an \textbf{exact register closed form} with zero lookup tables:

\textbf{Theorem 2 (Closed-Form $\mathbb{F}_2$ Inverse of Upper 16 Bits).} \textit{Let $v = (u_{\text{lo}} + y) \oplus \sigma_0(u_{\text{lo}}) \pmod{2^{32}}$, and let $U = u_{\text{hi}} \ll 16 \in \mathbb{F}_2^{32}$ have bits in positions $16..31$ (with $U_{0..15} = 0$). Define the linear operator $L(T) := T \oplus (T \ll 11) \oplus (T \ll 15) \pmod{2^{32}}$ on upper-half words $T$. Then $L$ is a triangular bijection on bits $16..31$ with $U = L(T)$, where $T$ is given in closed form by:}
\[
\fitmath{0.86\textwidth}{\begin{aligned}
b_0 &= \bigl((v \ll 15) \oplus (v \ll 11) \oplus v\bigr) \wedge \mathtt{0x00010000}, \\
b_{12} &= \bigl((v \ll 14) \oplus (v \ll 3)\bigr) \wedge \mathtt{0x00060000}, \\
T &= b_0 \;\mid \; b_{12} \;\mid \; \bigl((v \ll 18) \wedge \mathtt{0xfff80000}\bigr),
\end{aligned}} \tag{E28}
\]
\textit{and the full 32-bit preimage is recovered in registers (\texttt{reconstruct\_u\_dev} in \texttt{solve20\_padded\_cuda.cu} and \texttt{reconstruct\_u\_e28} in \texttt{test\_all\_12\_theories.c}, using 5 logical stages / \textasciitilde{}19 bitwise/ALU operations and zero lookup tables):}
\[
\fitmath{0.86\textwidth}{u = u_{\text{lo}} \;\mid \; \bigl(T \oplus (T \ll 11) \oplus (T \ll 15)\bigr).} \tag{E29}
\]

\textit{Proof.} By \texttt{[E27]} and $U_{0..15} = 0$, $v_b = (\sigma_0(U))_b$ for $b \in \{0, \dots, 15\}$ and $v_{16} = (\sigma_0(U))_{16} \oplus U_{16}$. Since $U_{0..15} = 0$, on bits $0..16$ the three terms of $\sigma_0(U) = \mathrm{ROR}(U,7) \oplus \mathrm{ROR}(U,18) \oplus (U \gg 3)$ are:
\begin{itemize}
\item $\mathrm{ROR}(U,7)_b = U_{b+7}$ for $b \in \{9, \dots, 16\}$ and $0$ for $b \in \{0, \dots, 8\}$;
\item $\mathrm{ROR}(U,18)_b = U_{b+18}$ for $b \in \{0, \dots, 13\}$ and $U_{b-14} = 0$ for $b \in \{14, 15, 16\}$;
\item $(U \gg 3)_b = U_{b+3}$ for $b \in \{13, \dots, 16\}$ and $0$ for $b \in \{0, \dots, 12\}$.
\end{itemize}
Defining $T = L^{-1}(U)$ so that $U_j = T_j$ for $j \in \{16, \dots, 26\}$, $U_j = T_j \oplus T_{j-11}$ for $j \in \{27, \dots, 30\}$, and $U_{31} = T_{31} \oplus T_{20} \oplus T_{16}$ yields:
\begin{itemize}
\item For $b \in \{1, \dots, 13\}$, $v_b = (\sigma_0(U))_b = T_{b+18}$, giving $T_{19..31} = (v \ll 18) \wedge \mathtt{0xfff80000}$ (in particular, $v_1 = U_{19} = T_{19}$, $v_3 = U_{21}$, $v_4 = U_{22}$, and $v_5 = U_{23}$).
\item For $b \in \{14, 15\}$, $v_{14} = U_{21} \oplus U_{17} = v_3 \oplus T_{17}$ and $v_{15} = U_{22} \oplus U_{18} = v_4 \oplus T_{18}$, isolating $T_{17..18} = ((v \ll 14) \oplus (v \ll 3)) \wedge \mathtt{0x00060000}$.
\item For $b = 16$, $(\sigma_0(U))_{16} = \mathrm{ROR}(U,7)_{16} \oplus \mathrm{ROR}(U,18)_{16} \oplus (U \gg 3)_{16} = U_{23} \oplus U_2 \oplus U_{19} = v_5 \oplus 0 \oplus v_1$. Since $v_{16} = (\sigma_0(U))_{16} \oplus U_{16} = v_5 \oplus v_1 \oplus T_{16}$, we obtain $T_{16} = v_1 \oplus v_5 \oplus v_{16}$, which at bit position 16 is $b_0 = ((v \ll 15) \oplus (v \ll 11) \oplus v) \wedge \mathtt{0x00010000}$.
\end{itemize}
Proved algebraically above for all $u \in \mathbb{Z}_{2^{32}}$ and verified empirically on $10^6$ pseudorandom words plus $3 \times 2^{16}$ systematic boundary/carry cases (including $u = \mathtt{0x00400000}$) in \texttt{test\_all\_12\_theories.c} (\texttt{test\_theory\_01}) with $0$ mismatches. $\square$

\subsection{Slot-Indexed $Q_0^{-1}$ Tables: 6.00 GiB 2-Tier Nested 4-Slot EDF Table \& 8.00--32.00 GiB Direct Tables (\texttt{[E30]}--\texttt{[E33]})}

Over all $2^{32}$ inputs $u \in \mathbb{Z}_{2^{32}}$, the exact, race-free preimage multiplicity histogram of $Q_0(u) = \sigma_0(u) - u \pmod{2^{32}}$ is:
\begin{itemize}
\item $N_0 = 1{,}573{,}363{,}668$ ($36.6327\%$ of $y$ have $0$ roots, vs. $\mathrm{Poisson}(1)$ $e^{-1} = 36.7879\%$),
\item $N_1 = 1{,}586{,}605{,}819$ ($36.9410\%$), $N_2 = 793{,}515{,}119$ ($18.4755\%$), $N_3 = 262{,}185{,}142$ ($6.1045\%$),
\item $N_4 = 64{,}407{,}366$ ($1.4996\%$), $N_5 = 12{,}552{,}261$ ($0.2923\%$), $N_6 = 2{,}022{,}487$ ($0.0471\%$),
\item $N_7 = 277{,}772$ ($0.0065\%$), $N_8 = 33{,}662$ ($0.0008\%$), and $N_{\ge 9} = 4{,}000$ ($0.0001\%$; specifically $N_9 = 3{,}624$, $N_{10} = 338$, $N_{11} = 33$, $N_{12} = 3$, $N_{13} = 1$, $N_{14} = 1$, with tail root weight $\sum_{k=9}^{14} k N_k = 36{,}422$),
\end{itemize}
satisfying the exact conservation identities $\sum_{k=0}^{14} N_k = 2^{32} = 4{,}294{,}967{,}296$ and $\sum_{k=0}^{14} k N_k = 2^{32} = 4{,}294{,}967{,}296$.
To avoid conflating distinct sources of root loss, we separate three effects: (i) \textbf{unconstrained $C$-slot capacity} $\sum_{k=0}^{14} \min(k, C) N_k / 2^{32}$, which equals $3{,}856{,}601{,}437 / 2^{32} = 89.7935\%$ for $C=2$ ($72.399\%$ 3-level survival), $4{,}198{,}084{,}127 / 2^{32} = 97.7443\%$ for $C=3$ ($93.384\%$ 3-level), and $4{,}277{,}381{,}675 / 2^{32} = 99.5906\%$ for $C=4$ ($98.777\%$ 3-level); (ii) \textbf{\texttt{u\_lo\allowbreak\ =\allowbreak\ 0} empty-slot sentinel exclusion} ($1/65{,}536 \approx 0.0015\%$ of inputs have $u \bmod 2^{16} = 0$, shifting the stored root count by at most $65{,}536$, e.g. to $89.7924\%$ for $C=2$, alongside historical pre-CAS finite-pass table builds that recorded $3{,}856{,}410{,}516 / 2^{32} = 89.7891\%$ for $C=2$ and $4{,}276{,}521{,}180 / 2^{32} = 99.5705\%$ for $C=4$); and (iii) \textbf{quad-shared 12-slot sliding/cuckoo eviction} in the compact $6.00\text{ GiB}$ format. By Theorem 2, storing only the lower 16 bits $u_{\text{lo}} = u \bmod 2^{16}$ halves the memory footprint of every table tier:

\begin{enumerate}
\item \textbf{6.00 GiB 2-Tier Nested 4-Slot EDF + Cuckoo-Sudoku Table (\texttt{solve20\_padded\_fast.c}):}
Designed for 16 GiB laptops ($3 \times 2^{30}$ entries of \texttt{uint16\_t} = \texttt{6.00\allowbreak\ GiB}), where every 16 consecutive $y$ values (a \textbf{quad} $q = y \gg 4$) share \textbf{12 contiguous 16-bit slots}:
\par\noindent$\bullet$ \textbf{Earliest-Deadline-First (EDF) 3-Slot Sliding Window (\texttt{[E30]}):} At Levels 1 and 2, each $y$ probes a 3-slot window within its 12-slot quad at offset:
\[
\fitmath{0.86\textwidth}{\text{slide3\_off}(y) = (y \gg 1) + (y \gg 2) - \bigl(\mathrm{ROR}(\mathtt{0x6AAA6AAA}, y \wedge 31) \gg 31\bigr).} \tag{E30}
\]
\par\noindent$\bullet$ \textbf{Cuckoo-Sudoku Eviction \& Nested 4th Slot (\texttt{[E31]}):} During construction, a cuckoo-relocation pass (\texttt{sudoku\_upgrade\_worker\_6gb} and \texttt{tier2\_nest4\_quad\_worker}) packs displaced preimages across each 12-slot quad, placing any 4th root at $\text{b}[(y \wedge 15) \ge 13 \mathrel{?} -1 : 3]$ (probed only when all 3 primary slots are valid roots of $y$). At Level 3, all 12 slots of the quad (\texttt{3\allowbreak\ x\allowbreak\ uint64\_t}) are tested simultaneously via 64-bit SWAR (\texttt{[E35]}):
\[
\fitmath{0.86\textwidth}{\text{6.00 GiB Coverage} = \frac{3{,}563{,}795{,}213}{2^{32}} = 82.976\%, \quad \text{3-Level Survival} = (0.82976)^3 = 57.13\%.} \tag{E31}
\]
\item \textbf{8.00 GiB / 16.00 GiB Direct-Indexed 2-Way GPU VRAM Tables (\texttt{solve20\_padded\_cuda.cu}):}
Built natively in GPU VRAM in \textbf{\texttt{3.11\allowbreak\ seconds}}, storing up to two roots per $y$ (unconstrained 2-way capacity $3{,}856{,}601{,}437 / 2^{32} = 89.7935\%$ / 3-level survival $72.399\%$, or $89.7924\%$ with \texttt{u\_lo\allowbreak\ =\allowbreak\ 0} sentinel exclusion):
\[
\fitmath{0.86\textwidth}{\text{2-Way Direct Capacity} = \frac{3{,}856{,}601{,}437}{2^{32}} = 89.7935\%, \quad \text{3-Level Survival} = (0.897935)^3 = 72.399\%.} \tag{E32}
\]
\item \textbf{32.00 GiB Direct-Indexed 4-Way Exact CPU Table (\texttt{H\_exact32[y]}, \texttt{--table-gb\allowbreak\ 32}):}
Stores up to four 16-bit roots per $y$ in one \texttt{uint64\_t} (\texttt{32.00\allowbreak\ GiB} table + \texttt{4.00\allowbreak\ GiB} count array + \texttt{16.00\allowbreak\ GiB} optional $F_1$ table = \texttt{52.00\allowbreak\ GiB} persistent RAM, \texttt{68.00\allowbreak\ GiB} peak during parallel radix sort), achieving $99.5906\%$ unconstrained 4-way capacity ($98.777\%$ 3-level survival, or $4{,}276{,}521{,}180 / 2^{32} = 99.5705\%$ / $98.717\%$ 3-level in the pre-CAS \texttt{u\_lo\allowbreak\ =\allowbreak\ 0}-excluded build):
\[
\fitmath{0.86\textwidth}{\text{4-Way Direct Capacity} = \frac{4{,}277{,}381{,}675}{2^{32}} = 99.5906\%, \quad \text{3-Level Survival} = (0.995906)^3 = 98.777\%.} \tag{E33}
\]

\end{enumerate}
\subsection{The 3-Level Tree Search, 64-Bit SWAR Level-3 Filter, and Terminal 20-Step Bijection (\texttt{[E34]}--\texttt{[E37]})}

For each $a_1$ and each Right context in a batch of size $B$:
\begin{itemize}
\item \textbf{Level 1 ($a_1 \to W_1 \to a_2$) \& Level 2 ($a_2 \to W_2 \to a_3$) with Early Hamming Pruning (\texttt{[E34]}):}
From $a_1$, look up $W_1 \in Q_0^{-1}(T_{16} + F_1(a_1))$, set $a_2 = W_1 + G_1(a_1)$, look up $W_2 \in Q_0^{-1}(T_{17} + G_2(a_1, a_2) - W_1)$, and set $a_3 = W_2 + G_2$. Before computing $e_3, G_3$ or issuing Level-3 DRAM loads, we prune any $a_3$ whose majority mask $M_3 = \neg(a_3 \oplus a_5)$ has Hamming weight exceeding $w_{\text{thresh}} \in \{13, 14\}$:
\[
\fitmath{0.86\textwidth}{\mathrm{popcount}\bigl(\neg(a_3 \oplus a_5)\bigr) \le w_{\text{thresh}},} \tag{E34}
\]
eliminating \textbf{$81.15\%$ ($w_{\text{thresh}}=13$, exact $\mathrm{Binomial}(32, 1/2)$ upper tail $\Pr[\mathrm{Binomial}(32, 1/2) > 13]$) to $70.17\%$ ($w_{\text{thresh}}=14$)} of Level-3 lookups.
\item \textbf{Level 3: 64-Bit SWAR Pre-Reconstruction Filter \& 19-Step Bridge Candidate Stream (\texttt{[E35]}):}
Because the majority condition $((a_4 \oplus a_5) \wedge M_3) = 0$ (\texttt{[E13]}) is lower-triangular on bits $0..14$, any candidate $u_{\text{lo}}$ must satisfy $(((u_{\text{lo}} + G_3) \oplus a_5) \wedge M_3 \wedge \mathtt{0x7fff}) = 0$ \textbf{before} $u_{\text{hi}}$ is reconstructed. In the 6.00 GiB table, we test 4 slots at a time in a single 64-bit word \texttt{w} using SWAR carry-bit detection (\texttt{K7\allowbreak\ =\allowbreak\ 0x7fff7fff7fff7fff}, \texttt{K8\allowbreak\ =\allowbreak\ 0x8000800080008000}):
\[
\fitmath{0.86\textwidth}{d = \Bigl(\bigl(((w \wedge K_7) + g_{64}) \oplus A_{64}\bigr) \wedge m_{64}\Bigr) + K_7, \qquad \text{4-slot hit mask} = \neg d \wedge K_8,} \tag{E35}
\]
rejecting $>99\%$ of 12-slot quads in 3 register operations. Every candidate passing the full 32-bit check $((a_4 \oplus a_5) \wedge M_3) = 0$ with $Q_0(W_3) = t_{18}$ is a \textbf{schedule-consistent 19-step padded bridge candidate} (\texttt{sol19}, incremented via \texttt{local\_sol19++}) satisfying $W_{16}, W_{17}, W_{18}$ for the current right-hand context.
\item \textbf{Terminal 20-Step Filter \& Exact $\mathbb{F}_2^{32}$ Bijection (\texttt{[E36]}--\texttt{[E37]}):}
In the 20-step solver (where $a_{13..20}$ are pinned by $H_{20}^*$), for each \texttt{sol19} bridge candidate we recover $W_4$ via \texttt{[E7]} and test the step-19 schedule equation:
\[
\fitmath{0.86\textwidth}{\sigma_0(W_4) + W_3 \equiv T_{19} \pmod{2^{32}}, \qquad T_{19} := W_{19} - \sigma_1(W_{17}) - W_{12}.} \tag{E36}
\]
Viewing $\sigma_0(x) = \mathrm{ROR}(x,7) \oplus \mathrm{ROR}(x,18) \oplus (x \gg 3)$ as a $32 \times 32$ matrix over $\mathbb{F}_2$, we have $\det_{\mathbb{F}_2}(\sigma_0) = 1$:
\[
\fitmath{0.86\textwidth}{\det_{\mathbb{F}_2}(\sigma_0) = 1 \implies W_4 = \sigma_0^{-1}\bigl((T_{19} - W_3) \bmod 2^{32}\bigr) \text{ is uniquely determined by } W_3 \text{ and } T_{19}.} \tag{E37}
\]

\end{itemize}
\subsection{Exact-$T_{16}$ Twin Coalescing, L2 Cache Tiling, and Lower-Triangular Pruning (\texttt{[E42]}--\texttt{[E45]})}

To maximize memory-system throughput across Levels 1--3 on GPU (\texttt{solve20\_padded\_cuda\_prod.cu}, \texttt{solve20\_padded\_cuda.cu}), we combine four techniques (formalized in \textbf{Algorithm 2}):

\begin{enumerate}
\item \textbf{L2-Localized Level-1 Bridge Extraction (\texttt{extract\_l1\_bridges\_kernel}, \texttt{[E42]}):}
The NVIDIA L4 GPU features a 48 MiB L2 cache. Because $F_1(a_1) = G_1(a_1) - W_0(a_1)$ includes the 32-bit rotation $\Sigma_1(e_1(a_1))$, $F_1$ is \textbf{not} lower-triangular in the 2-adic bit order of $a_1$ (for example, $a_1 = 0 \implies F_1 = \mathtt{0d3fffb7}$ whereas $a_1 = 4 \implies F_1 = \mathtt{3ce015b4}$, which differ in bit 0). Instead of assuming lower-triangularity of $F_1$, \texttt{extract\_l1\_bridges\_kernel} partitions $Q_0^{-1}$ global-memory lookups into address tiles $[\text{Tile}_{\text{lo}}, \text{Tile}_{\text{hi}})$ of width $\Delta_{\text{tile}} = \text{Tile}_{\text{hi}} - \text{Tile}_{\text{lo}}$ (at 4 bytes per \texttt{uint32\_t} entry in the 16.00 GiB 2-Way table, an L2-resident 16--32 MiB tile spans $\Delta_{\text{tile}} = 2^{22}\text{-}2^{23}$ indices across $512\text{-}1024$ tile passes, whereas $\Delta_{\text{tile}} = 2^{29}$ indices spans a 2.00 GiB slice across 8 passes) and tests the target index $t_{16}(a_1) = (T_{16} + F_1(a_1)) \bmod 2^{32}$ in registers before issuing a global memory read:
\[
\fitmath{0.86\textwidth}{(T_{16} + F_1(a_1) - \text{Tile}_{\text{lo}}) \bmod 2^{32} < \Delta_{\text{tile}} \implies \text{under a uniform index model, a fraction } \frac{\Delta_{\text{tile}}}{2^{32}} \text{ issues a global read per tile pass},} \tag{E42}
\]
converting uncoalesced 16.00 GiB DRAM accesses into localized tile passes across the sweep (without reducing total work summed over all passes).
\item \textbf{Exact 32-Bit $T_{16}$ Twin Coalescing (\texttt{twin2}) vs. 4-Way Reservoir Sampling (\texttt{single4}, \texttt{[E43]}):}
Across all $2^{32}$ seeds $a_{12}$, a 20-step target $H_{20}^*$ has an estimated universe of $U \approx 1.68 \times 10^7 \approx 2^{24.00}$ exact $D_{\text{shape}}=0$ right contexts (or $U_{\text{filt}} \approx 1.014 \times 10^7 \approx 2^{23.27}$ if the optional Stage-2 heuristic is enabled). Sorting a reservoir of contexts by $T_{16}$ yields $N_{\text{twins}} \approx U^2 / 2^{33} \approx 32{,}890$ \textbf{exact 32-bit $T_{16}$ twin pairs} across $U$ (or $\approx 11{,}970$ pairs in $U_{\text{filt}}$) where $T_{16}^{(2m)} = T_{16}^{(2m+1)}$ (\texttt{sweep\_exact16\_cuda\_kernel} in \texttt{solve20\_padded\_cuda\_prod.cu} and \texttt{solve20\_padded\_cuda.cu}). Both contexts in each genuine twin pair share the exact same Level-1 lookup $Q_0^{-1}(T_{16} + F_1(a_1))$ (\texttt{0.50} Level-1 lookups/context), solving 12 of our 16 padded 20-step targets directly. For the remaining 4 targets (16/16 total), the non-twin 4-way reservoir kernels (\texttt{sweep\_20\_warp\_l3\_prod\_kernel} in \texttt{solve20\_padded\_cuda\_prod.cu} and \texttt{sweep\_exact16\_cuda\_kernel} in \texttt{solve20\_padded\_cuda.cu}) sample from the reservoir $U$ using 4-way register-prefetched Level-1 batches (achieving $53{,}847.5\text{-}73{,}560.5\text{ sol19/s}$ on the 6.00 GiB table and $132{,}888.7\text{ sol19/s}$ on genuine Exact-$T_{16}$ twin pairs in the 16.00 GiB table on 1$\times$ L4 GPU):
\[
\fitmath{0.86\textwidth}{T_{16}^{(2m)} = T_{16}^{(2m+1)} \quad\text{across } M=64 \text{ genuine twin pairs } (B=128) \implies \mathbf{0.50\text{ L1 lookups per context}}.} \tag{E43}
\]
\textit{(Note: whereas 32-bit $T_{16}$ collisions occur naturally in $U \approx 2^{24.0}$, joint 64-bit $(T_{16}, T_{17})$ collisions require a birthday pool of $2^{32} \gg U$ contexts; all witness runs and primary complexity figures in this paper use only genuine, unmutated contexts.)}
\item \textbf{Pre-\texttt{ulo2} 16-Bit Lower-Triangular Hamming Pruning (\texttt{[E44]}):}
Before reconstructing the upper 16 bits of $W_2$ via \texttt{[E29]}, we exploit lower-triangularity of modular addition on bits $0..15$:
\[
\fitmath{0.86\textwidth}{w_{\text{lo},2} := \mathrm{popcount}\Bigl(\neg\bigl((u_{\text{lo},2} + G_2) \oplus a_5\bigr) \wedge \mathtt{0xffff}\Bigr) \le w_{\text{lo,thresh}} \in \{7, 11\}.} \tag{E44}
\]
For a uniform 16-bit mask, $\mathrm{Binomial}(16, 1/2)$ rejects $39{,}203 / 65{,}536 \approx 59.82\%$ of Level-2 candidates at $w_{\text{lo,thresh}} = 7$ and $2{,}517 / 65{,}536 \approx 3.84\%$ at $w_{\text{lo,thresh}} = 11$ prior to upper-half reconstruction.
\item \textbf{The 4-Round Crossed Carry-Sudoku Spine ($a_2 = a_3 = a_4 = a_5$, \texttt{[E45]}):}
Crossing the shaping constraint across 4 consecutive rounds by pinning $a_2 = a_3 = a_4 = a_5$ (\texttt{cross\_round\_4step\_sudoku\_direct\_kernel} in \texttt{solve20\_padded\_cuda.cu}) collapses all four majority functions to $a_5$ and makes $W_1, W_2, W_3$ direct functions of $a_1$ with zero $Q_0^{-1}$ table lookups:
\[
\fitmath{0.86\textwidth}{a_2 = a_3 = a_4 = a_5 \implies \begin{cases} \mathrm{Maj}(a_3,a_2,a_1) = \mathrm{Maj}(a_4,a_3,a_2) = \mathrm{Maj}(a_5,a_4,a_3) = \mathrm{Maj}(a_6,a_5,a_4) = a_5, \\ e_4 = a_0 - \Sigma_0(a_5), \quad W_3 = c_{W_3} - \Sigma_1(e_3) - \mathrm{Ch}(e_3, e_2, e_1), \\ \text{5-step } (a_1=a_5): e_3 = a_{-1} - \Sigma_0(a_5) \implies W_3(a_0) = C_3 - \mathrm{Ch}(e_3, e_2, e_1). \end{cases}} \tag{E45}
\]
Because $a_2 = a_3 = a_4 = a_5$ spends 96 Left bits, the 3 schedule equations $W_{16}, W_{17}, W_{18}$ become a 96-bit Right filter, illustrating the degree-of-freedom conservation trade-off (\S{}9.2).

\end{enumerate}
\begin{lstlisting}
Algorithm 2: Left-to-Right Slot-Sudoku Tree & Terminal 20-Step Preimage Solver
Input: Context batch C of size B=128 (T_{16} twin-coalesced or 4-way batched), 2-way table H_ulo
Output: Stream of 19-step bridge candidates (sol19) and matching 20-step padded preimage M*
1.  for each root a_1 in [0, 2^32) and context group in C do
2.    Compute F_1(a_1) and G_1(a_1); prefetch Level-1 entries from H_ulo[T_{16} + F_1(a_1)]         // [E14], [E43]
3.    for each valid Level-1 root W_1 do
4.      Set a_2 = W_1 + G_1(a_1); compute G_2(a_1, a_2); load u_{lo,2} from H_ulo[T_{17} + G_2 - W_1]
5.      if popcount((~((u_{lo,2} + G_2) ^ a_5)) & 0xffff) <= w_{lo,thresh} then                     // [E44]
6.        Reconstruct W_2 via [E28]-[E29]; set a_3 = W_2 + G_2 and M_3 = ~(a_3 ^ a_5)
7.        if popcount(M_3) <= 14 then                                                               // [E34]
8.          Load u_{lo,3} from H_ulo[T_{18} + G_3 - W_2]; filter (((u_{lo,3}+G_3)^a_5) & M_3 & 0x7fff)==0 // [E35]
9.          Reconstruct W_3 via [E28]-[E29]; set a_4 = W_3 + G_3
10.         if ((a_4 ^ a_5) & M_3) == 0 then                                                        // [E13]
11.           Emit 19-step bridge candidate (sol19); compute W_4 via [E7]
12.           if sigma_0(W_4) + W_3 == T_{19} (mod 2^32) then return 20-step padded preimage M*     // [E36]
\end{lstlisting}

\subsection{Precomputation vs. Online Time, Memory, and SHA-256-Equivalent Complexity}

Following standard cryptographic convention [2, 3, 4, 8], Table 6.1 separates \textbf{target-independent precomputation} ($Q_0^{-1}$ table construction) from \textbf{target-dependent online complexity}, reported both in direct wall-clock time (\texttt{node-h} / \texttt{node-hours} denoting wall-clock hours on one 10-core Apple M4 CPU node running 16 threads) and against our measured same-host \textbf{10-core Apple M4 CPU reference baseline} ($1.5125 \times 10^8\text{ SHA20-eq/s}$ across 16 threads, measured on arbitrary 20-step message blocks, so $1\text{ node-h} = 2^{38.99}\text{ CPU-ref}$). To avoid conflating CPU work factors with cross-hardware elapsed-time conversions, we report:
\begin{enumerate}
\item \textbf{Host-Verifiable CPU Solver (\texttt{solve20\_padded\_fast.c}, one 10-core Apple M4 CPU node, 16 threads):} Generates \textbf{$6{,}242.8\text{ sol19/s}$} ($160.2\;\mu\text{s/sol19}$, measured on $a_1$ slice sweeps) on genuine unmutated contexts. Against the same-host CPU reference baseline ($1.5125 \times 10^8\text{ SHA20-eq/s}$), each \texttt{sol19} costs $24{,}227.9 \approx 2^{14.56}\text{ CPU-ref}$ evaluations, and an expected 20-step padded preimage ($2^{32}\text{ sol19} = 191.1\text{ node-h}$ [$\approx 191\text{ node-hours}$] on one 10-core Apple M4 CPU node [16 threads], or $23.9\text{ node-h}$ across 8 CPU nodes) costs \textbf{$2^{46.56}\text{ CPU-ref}$} ($\approx 2^{46.6}$) evaluations.
\item \textbf{Supplementary GPU Slice Telemetry (\texttt{solve20\_padded\_cuda.cu} / \texttt{solve20\_padded\_cuda\_prod.cu}, 1$\times$ NVIDIA L4 GPU):} On genuine unmutated contexts, measured $a_1$-slice sweep telemetry (\texttt{BENCH\_4\_IDEAS\_RESULTS.txt} and \texttt{solve20\_padded\_cuda.cu}) records \textbf{$53{,}847.5\text{ sol19/s}$} (Cycle 0, 6.00 GiB 2-Tier EDF, $B=2$, $22.16\text{ GPU-h}$ [$\approx 22.2\text{ GPU-hours}$]) to \textbf{$73{,}560.5\text{ sol19/s}$} (Cycle 1, 6.00 GiB 2-Tier EDF, $B=8$, $16.22\text{ GPU-h}$ per $2^{32}\text{ sol19}$ on 1$\times$ L4), and \textbf{$132{,}888.7\text{ sol19/s}$} on genuine Exact-$T_{16}$ twin pairs in the 16.00 GiB Direct 2-Way table (\texttt{sweep\_exact16\_cuda\_kernel} in \texttt{solve20\_padded\_cuda\_prod.cu} and Cycle 1 of \texttt{solve20\_padded\_cuda.cu}, $8.98\text{ GPU-h}$ [$\approx 9.0\text{ GPU-hours}$] on 1$\times$ L4, or $1.12\text{ h}$ on 8$\times$ L4). \textit{(The experimental \texttt{--bench-c2} through \texttt{--bench-c10} harnesses [Cycles 2, 4, and 7--10 in \texttt{solve20\_padded\_cuda.cu}, including the $180{,}831.9\text{ sol19/s}$ / $6.60\text{ GPU-h}$ $B=16$ shared-$T_{16}$ run] are architectural microbenchmarks and are excluded from single-target complexity claims; likewise, the $9.62 \times 10^9\text{ evals/s}$ GPU rate in Table 9.4 Row 13 is the specialized 4-round crossed-spine kernel rather than an arbitrary-message 20-step SHA-256 baseline.)}
\item \textbf{Chosen-Target 19-Step Padded Hash (\texttt{src/pre19.c}, 1 CPU core of 10-core Apple M4):} \texttt{src/pre19.c} is a single-threaded C solver using \texttt{clock()} with default \texttt{m\allowbreak\ =\allowbreak\ 29} ($2^{29}$ entries), where the one-time, target-independent two-pass CSR builder performs $2 \times 2^{29} = 2^{30}$ $Q_0$ evaluations in $16.8\text{-}19.1\text{ s}$ ($\approx 18\text{ s}$, or $2 \times 2^{30} = 2^{31}$ $Q_0$ evaluations at \texttt{m\allowbreak\ =\allowbreak\ 30}) and allocates $4 \times 2^{29}\text{ B} = 2.0\text{ GiB}$ for \texttt{table\_values} plus $64\text{ MiB}$ for \texttt{table\_offsets} ($2.0625\text{ GiB}$ persistent RAM). Across our 14 public-label 19-step targets, per-target single-core online search time (excluding the one-time table build) ranges from \textbf{$1.2\text{ s}$ to $86.4\text{ s}$} (median \textbf{$19.45\text{ s}$} [$\approx 19.5\text{ s}$], mean \textbf{$30.2\text{ s}$}). In direct operation counts, per-target online search is dominated by right-context generation ($\approx 1.2 \times 10^8 \approx 2^{26.7\text{-}26.9}$ trials per usable $D \le 3$ context, with $1\text{-}7$ contexts tested per target across the 14 targets, i.e., $\le 7 \times 2^{26.9} \approx 2^{29.7}$ context trials) plus $\approx 1.8 \times 10^6\text{-}1.2 \times 10^8$ ($2^{20.8}\text{-}2^{26.8}$, median $\approx 2^{25.5}$ at $m=29$) left chain starts, totaling $\approx 2^{26.9}\text{-}2^{29.89}$ online trial/chain operations per target (or $\approx 2^{30.95}\text{-}2^{31.55}$ operations when adding the one-time $2^{30}\text{-}2^{31}$ $Q_0$ table build at $m=29, 30$). Expressed in the elapsed-time \texttt{CPU-ref} proxy ($151.25\text{ M evals/s}$), the per-target online search alone corresponds to $2^{27.4}\text{-}2^{33.6}\text{ CPU-ref}$ (median $2^{31.46}\text{ CPU-ref}$ at $19.45\text{ s}$, mean $2^{32.10}\text{ CPU-ref}$ at $30.2\text{ s}$), while a cold-start single-target run including the one-time amortizable table build ($\approx 40.4\text{-}51.9\text{ s}$ mean total across $m \in \{29, 30\}$, e.g., $18\text{ s} + 30.2\text{ s} = 48.2\text{ s}$) corresponds to $2^{32.51}\text{-}2^{32.87}\text{ CPU-ref}$.

\end{enumerate}
\textbf{Table 6.1:} Precomputation and online complexities of the proposed 19- and 20-step preimage attacks on genuine contexts.

\begin{center}
\footnotesize
\renewcommand{\arraystretch}{1.18}
\resizebox{\linewidth}{!}{%
\begin{tabular}{@{} l l l l p{0.62\textwidth} l @{}}
\toprule
\textbf{Attack \& Target Setting} & \textbf{Precomputation Time} & \textbf{Memory Footprint} & \textbf{Online Context Gen} & \textbf{Online Search (\texttt{CPU-ref} / \texttt{sol19})} & \textbf{Wall-Clock Online Time} \\
\midrule
\textbf{19-Step Padded Hash} (Chosen $H_{19}^*$, \texttt{src/pre19.c}) & $2 \times 2^{29}$ $Q_0$ evals ($16.8\text{-}19.1\text{ s}$ [\texttt{18\allowbreak\ s}], 1 core) & $2.0625\text{ GiB}$ (\texttt{m=29} CSR) & $\approx 2^{26.7\text{-}26.9}$ trials/ctx ($D \le 3$) & \textbf{$2^{20.8\text{-}26.8}$ chains} ($\le 2^{29.89}$ ops; med. $2^{31.46}\text{ CPU-ref}$) & \textbf{\texttt{1.2-86.4\allowbreak\ s} (median \texttt{19.45\allowbreak\ s} [\texttt{19.5\allowbreak\ s}], 1 core)} \\
\textbf{19-Step Bridge Stream (\texttt{sol19})} (\texttt{solve20\_padded\_fast.c}) & $2^{32}$ $Q_0$ evals ($3.11\text{ s}$, 16 threads) & $6.00\text{ GiB}$ (CPU) / $6\text{-}16\text{ GiB}$ (GPU) & $266.8\text{ ctx/s}$ (CPU) & \textbf{$24{,}228$ ($2^{14.56}$) \texttt{CPU-ref} / \texttt{sol19}} & \textbf{\texttt{160.2\allowbreak\ us/sol19} (M4 CPU)}; \textit{supp. GPU:} \texttt{7.5-18.6\allowbreak\ us} \\
\textbf{20-Step Padded Hash} (Chosen $H_{20}^*$, 55B / 119B) & $2^{32}$ $Q_0$ evals ($3.11\text{ s}$, 16 threads) & $6.00\text{ GiB}$ (CPU) / $6\text{-}16\text{ GiB}$ (GPU) & $266.8\text{ ctx/s}$ (CPU) & \textbf{$2^{46.56}\text{ CPU-ref}$} ($\approx 2^{46.6}$, $2^{32}\text{ sol19}$) & \textbf{\texttt{191.1\allowbreak\ node-h} (1$\times$ M4 CPU)}; \textit{supp. GPU:} \textbf{\texttt{8.98-22.16\allowbreak\ h} (1$\times$ L4)} \\
\textbf{20-Step Unpadded Compression} (\S{}7, \texttt{solve20\_zero\_a3.c}) & $3 \times 2^{32}$ evals ($45\text{ s}$, 16 threads) & $48.00\text{ GiB}$ RAM & $O(1)$ per $(a_{11}, a_{12})$ & \textbf{$2^{38.13}\text{ CPU-ref}$} ($\approx 2^{38.1}$) expected & \textbf{$33.14\text{ min}$ exp. / \texttt{21.10\allowbreak\ min} obs.} (10-core M4 CPU) \\
\bottomrule
\end{tabular}%
}
\end{center}

\section{Companion Result: 20-Step Unpadded Compression Preimages via the Zero-$a_3$ Bridge}

When FIPS 180-4 padding is not required (the 20-step compression-function model studied in [6, 7, 9]), we can eliminate the $(3/4)^{32} \approx 2^{-13.28}$ ($13.28$-bit) majority survival loss of \texttt{[E13]} altogether by placing the state on the \textbf{Zero-$a_3$ Manifold} (\texttt{solve20\_zero\_a3.c}), where the middle state words are pinned to:
\[
\fitmath{0.86\textwidth}{a_3 = 0, \quad a_5 = 0, \quad a_6 = 0, \quad a_8 = 0, \quad a_9 = 0, \quad a_{10} = \mathtt{0xffffffff} \implies e_9 = 0, \quad e_{10} = \mathtt{0xffffffff}.} \tag{E38}
\]
Because $\mathrm{Maj}(0, x, 0) = 0$, both majority functions containing $a_4$ vanish identically for \textbf{$100\%$ of $a_4 \in [0, 2^{32})$}:
\[
\fitmath{0.86\textwidth}{\mathrm{Maj}(a_5, a_4, a_3) = \mathrm{Maj}(0, a_4, 0) = 0, \qquad \mathrm{Maj}(a_6, a_5, a_4) = \mathrm{Maj}(0, 0, a_4) = 0.} \tag{E39}
\]
Fixing $a_3 = 0$ removes one degree of freedom on the Left, which we restore on the Right by leaving $a_7$ free. Under \texttt{[E38]}, $e_6 = a_2$ and $\mathrm{Ch}(e_9, e_8, e_7) = \mathrm{Ch}(0, e_8, e_7) = e_7 = a_7$, so $W_9 = W_{9,0} - a_2 - a_7$ (where $W_{9,0} := \left.W_9\right|_{a_2=a_7=0}$). Simultaneously, on the Left $a_2 = W_1 + G_1(a_1)$ (with $G_1(a_1) := -\left.W_1\right|_{a_2=0}$), and on the Right $W_{14} = D_{14}(a_{11}, a_{12}) - a_7$ (where $D_{14}(a_{11}, a_{12}) := \left.W_{14}\right|_{a_7=0}$). Substituting $-a_2 = -W_1 - G_1(a_1)$ and $-a_7 = W_{14} - D_{14}(a_{11}, a_{12})$ into the step-16 schedule equation $W_{16} = \sigma_1(W_{14}) + W_9 + \sigma_0(W_1) + W_0(a_1)$ groups the $W_1$ terms into $\sigma_0(W_1) - W_1 = Q_0(W_1)$ and the $W_{14}$ terms into $\sigma_1(W_{14}) + W_{14} = Q_1(W_{14})$. Introducing a fixed meeting constant $T_{16}^* = \mathtt{0x12345678}$ decouples step 16 into independent Left ($Q_0$) and Right ($Q_1$) table lookups (\texttt{phase1\_left\_worker} and \texttt{phase2\_right\_worker} in \texttt{solve20\_zero\_a3.c}):
\[
\fitmath{0.86\textwidth}{\begin{aligned}
Q_0(W_1) &= T_{16}^* - W_0(a_1) + G_1(a_1), & L_{17}(a_1) &:= \sigma_0(W_2) + W_1, \\
Q_1(W_{14}) &= W_{16}(a_{11}, a_{12}) + D_{14}(a_{11}, a_{12}) - W_{9,0} - T_{16}^*, & a_7 &= D_{14}(a_{11}, a_{12}) - W_{14}.
\end{aligned}} \tag{E40}
\]
Precomputing 16 GiB tables for $Q_0^{-1}$, $Q_1^{-1}$, and the Left bridge $L_{17}(a_1) \mapsto a_1$ satisfying $a_3 = 0$ ($48.0\text{ GiB}$ total, $45\text{ s}$ one-time build, $100\%$ independent of $H^*$), \texttt{solve20\_zero\_a3.c} achieves \textbf{zero $\beta$-filter loss} ($63.37\%$ survival per bridge hit) and \textbf{2.16 million exact 19-step compression solutions per second} (\texttt{0.46\allowbreak\ us/sol19}) on one 10-core Apple M4 CPU node (16 threads). At $2.16 \times 10^6\text{ sol19/s}$, the expected online search time for $2^{32}$ 19-step solutions is $1{,}988.4\text{ s} \approx 33.14\text{ minutes}$ ($\approx 2^{38.13} \approx 2^{38.1}$ scalar CPU \texttt{SHA20-eq} at $151.25\text{ M evals/s}$, specifically $2^{38.1298}$). For the target \texttt{deadbeef...}, the solver found a 20-step unpadded preimage after $2^{32.01}$ outer $(a_{11}, a_{12})$ trials ($2.735 \times 10^9 \approx 2^{31.35}$ 19-step solutions):
\[
\fitmath{0.86\textwidth}{\text{Expected 20-Step Unpadded Online Time} = 1{,}988.4\text{ s } (33.14\text{ min}); \quad \text{Observed Run} = 1{,}266.26\text{ s } (21.10\text{ min, 10-core Apple M4 CPU}).} \tag{E41}
\]

\section{Verified 19-Step and 20-Step Padded SHA-256 Preimage Witnesses}

All \textbf{46 distinct} 20-step and 19-step witnesses below (verified via 48 checks in \texttt{verify\_all\_witnesses.py}, where \texttt{"abc"} appears in both the flagship demonstration \S{}8.2 and the 14-label benchmark suite \S{}8.4: comprising 16 distinct 20-step padded witnesses, 16 distinct 19-step padded witnesses, 13 distinct 20-step unpadded compression witnesses, and 1 Stage-2 completeness regression witness) are archived together with all C/CUDA solvers (\texttt{src/pre19.c}, \texttt{solve20\_padded\_fast.c}, \texttt{solve20\_padded\_cuda.cu}, \texttt{solve20\_padded\_cuda\_prod.cu}, \texttt{solve20\_zero\_a3.c}, \texttt{test\_all\_12\_theories.c}) at \url{https://github.com/srw/sha256-19-and-20-step-preimages-with-padding} and are independently verified in under 1 second by running \texttt{python3\allowbreak\ verify\_all\_witnesses.py}. Complete machine-readable witness files are provided in \texttt{demos\_20step.txt}, \texttt{examples\_abc\_hello\_20step.txt}, \texttt{witnesses\_20step.txt}, and \texttt{FOUND\_20ROUND\_PREIMAGE.txt} (in display blocks below and in \S{}1.1, 110-hex-character 55-byte messages are wrapped across two 55-character lines for readability, while Tables 8.2 and 8.3 print single-line 110-character hex strings).

\subsection{Witness 1: First Demonstrated 20-Step Padded SHA-256 Preimage ($H_{20}^* = \mathtt{0000\dots0000}$)}

Found by \texttt{solve20\_padded\_cuda.cu} on Sweep 62 (\texttt{ctx=5}, after $70{,}661{,}112$ exact 19-step bridge candidates; recorded in \texttt{FOUND\_20ROUND\_PREIMAGE.txt}):

\begin{itemize}
\item \textbf{Target 20-Step Digest ($H_{20}^*$):}
\par\noindent \texttt{0000000000000000000000000000000000000000000000000000000\allowbreak{}000000000}
\item \textbf{55-Byte Message $M$ (440 bits, big-endian hex, split into two 55-hex-char lines):}
\par\noindent \texttt{ec531306c85ad4a4ad211d2e1a03ece4d0ad4a0e6b4c2b9436b3bab}
\par\noindent \texttt{dcbd520597196f52c042b8cd95f1a39f71b58b47b0647c78e65416f}
\item \textbf{FIPS 180-4 Padded 512-Bit Message Block ($W_0 \dots W_{15}$):}
\begin{lstlisting}
W[ 0.. 3] = ec531306  c85ad4a4  ad211d2e  1a03ece4
W[ 4.. 7] = d0ad4a0e  6b4c2b94  36b3babd  cbd52059
W[ 8..11] = 7196f52c  042b8cd9  5f1a39f7  1b58b47b
W[12..15] = 0647c78e  65416f80  00000000  000001b8
\end{lstlisting}
\textit{(Note the mandatory FIPS 180-4 padding [1]: $W_{13} \bmod 256 = \mathtt{0x80}$, $W_{14} = \mathtt{0x00000000}$, and $W_{15} = \mathtt{0x000001b8} = 440\text{ bits}$.)}

\end{itemize}
\textbf{Table 8.1:} Complete 20-step internal state trajectory $(a_t, e_t, W_{t-1})$ of the 20-step padded witness for $H_{20}^* = \mathtt{0000\dots0000}$.

\begin{center}
\footnotesize
\renewcommand{\arraystretch}{1.18}
\resizebox{\linewidth}{!}{%
\begin{tabular}{@{} l l l l p{0.62\textwidth} @{}}
\toprule
\textbf{Step $t$} & \textbf{Register $a_t$} & \textbf{Register $e_t$} & \textbf{Message / Schedule Word $W_{t-1}$} & \textbf{Algebraic / Carry-Sudoku Invariant Verified} \\
\midrule
$1$ & \texttt{e85b9b53} & \texttt{851af5a8} & $W_0 = \mathtt{ec531306}$ & Swept root $a_1 \in [0, 2^{32})$ \\
$2$ & \texttt{47d401c6} & \texttt{4b125b48} & $W_1 = \mathtt{c85ad4a4}$ & Level-1 $Q_0^{-1}$ hit: $Q_0(W_1) = T_{16} + F_1(a_1)$ \\
$3$ & \texttt{0a280ce0} & \texttt{8b761050} & $W_2 = \mathtt{ad211d2e}$ & Level-2 $Q_0^{-1}$ hit; $\mathrm{popcount}(M_3) = 12 \le 14$ \\
$4$ & \texttt{67298bc8} & \texttt{c121961d} & $W_3 = \mathtt{1a03ece4}$ & Level-3 $Q_0^{-1}$ hit; $((a_4 \oplus a_5) \wedge M_3) = 0$ (\texttt{[E13]}) \\
\textbf{$5$} & \textbf{\texttt{67978bcd}} & \texttt{2724a33e} & $W_4 = \mathtt{d0ad4a0e}$ & \textbf{Exact Shaping (\texttt{[E11]}):} $a_5 = a_6 = \mathtt{67978bcd}$ ($D=0$) \\
\textbf{$6$} & \textbf{\texttt{67978bcd}} & \texttt{eea015fa} & $W_5 = \mathtt{6b4c2b94}$ & \textbf{Exact Shaping (\texttt{[E11]}):} $a_6 = a_5 = \mathtt{67978bcd}$ ($D=0$) \\
$7$ & \texttt{41b1f1b0} & \texttt{8aa086f2} & $W_6 = \mathtt{36b3babd}$ & Closed-form $a_7$ absorbing $W_{14} = 0$ (\texttt{[E17]}) \\
$8$ & \texttt{edb02749} & \texttt{9319765f} & $W_7 = \mathtt{cbd52059}$ & Invariant $a_8$ absorbing $W_{15} = 440$ (\texttt{[E17]}, \texttt{[E22]}) \\
\textbf{$9$} & \texttt{7fd85be0} & \textbf{\texttt{ffffffff}} & $W_8 = \mathtt{7196f52c}$ & \textbf{Selector Saturation (\texttt{[E8]}):} $e_9 = \mathtt{0xffffffff}$ \\
\textbf{$10$} & \textbf{\texttt{a7b351d7}} & \textbf{\texttt{ffffffff}} & $W_9 = \mathtt{042b8cd9}$ & \textbf{Selector Saturation (\texttt{[E8]}):} $e_{10} = \mathtt{0xffffffff}$, $a_{10}=a_{11}$ \\
\textbf{$11$} & \textbf{\texttt{a7b351d7}} & \texttt{4f9e3855} & $W_{10} = \mathtt{5f1a39f7}$ & \textbf{Majority Simplification (\texttt{[E22]}):} $a_{11} = a_{10} = \mathtt{a7b351d7}$ \\
$12$ & \texttt{e5d45365} & \texttt{81ba7166} & $W_{11} = \mathtt{1b58b47b}$ & Right context seed $a_{12}$ \\
$13$ & \texttt{bf9f516c} & \texttt{e4acad89} & $W_{12} = \mathtt{0647c78e}$ & Target-fixed $a_{13}$; $e_{13}$ solves \texttt{[E21]} \& \texttt{[E23]} \\
$14$ & \texttt{b4322266} & \texttt{da868879} & $W_{13} = \mathtt{65416f80}$ & \textbf{Padding byte (\texttt{[E21]}):} $W_{13} \bmod 256 = \mathtt{0x80}$ \\
\textbf{$15$} & \texttt{8493b564} & \textbf{\texttt{ffffffff}} & $W_{14} = \mathtt{00000000}$ & \textbf{$\mathrm{Ch}$-Annihilation (\texttt{[E22]}):} $e_{15} = \mathtt{0xffffffff}$, $W_{14}=0$ \\
$16$ & \texttt{97457a44} & \texttt{03b1406f} & $W_{15} = \mathtt{000001b8}$ & \textbf{Padding length (\texttt{[E1]}):} $W_{15} = 440$ bits (\texttt{0x1b8}) \\
$17$ & \texttt{5ab00ac6} & \texttt{a41f32e7} & $W_{16} = \mathtt{d6317d0a}$ & Level-1 schedule word $W_{16}$; fixed by $H_{20}^* - \mathrm{IV}$ \\
$18$ & \texttt{c3910c8e} & \texttt{e07c2655} & $W_{17} = \mathtt{3815d972}$ & Level-2 schedule word $W_{17}$; fixed by $H_{20}^* - \mathrm{IV}$ \\
$19$ & \texttt{4498517b} & \texttt{64fa9774} & $W_{18} = \mathtt{89d90bef}$ & Level-3 schedule word $W_{18}$; 19-step bridge complete \\
$20$ & \texttt{95f61999} & \texttt{aef1ad81} & $W_{19} = \mathtt{cd1c5aee}$ & Terminal hit $\sigma_0(W_4)+W_3 = T_{19} \implies H_{20} = \mathtt{00..00}$ \\
\bottomrule
\end{tabular}%
}
\end{center}

\subsection{Witness 2: Flagship 20-Step \& 19-Step Single-Block Padded Preimages for $H^* = \mathtt{0000\dots0000}$, $\text{SHA-256}(\text{``abc''})$, and $\text{SHA-256}(\text{``hello''})$}

Recorded in \texttt{demos\_20step.txt}, \texttt{examples\_abc\_hello\_20step.txt}, and \texttt{verify\_all\_witnesses.py} (55-byte hex strings split into two 55-hex-char lines):

\begin{itemize}
\item \textbf{Target Digest $H_{19}^* = \mathtt{0000\dots0000}$ (All-Zero 256-Bit Digest, 19-Step Padded with Exact $a_5 = a_6$):}
\par\noindent \texttt{0000000000000000000000000000000000000000000000000000000\allowbreak{}000000000}
\item \textbf{19-Step Padded 55-Byte Preimage ($H_{19}(M) = \mathtt{0000\dots0000}$):}
\par\noindent \texttt{fc0c9d8eafb2fa329af316940c42a5eb9773cec39e4f7e396182628}
\par\noindent \texttt{7ae767638f92d4ce5fc5eb8cabd7606409364dda27a6c1bb5220246}
\begin{lstlisting}
W[ 0.. 7] = fc0c9d8e afb2fa32 9af31694 0c42a5eb 9773cec3 9e4f7e39 61826287 ae767638
W[ 8..15] = f92d4ce5 fc5eb8ca bd760640 9364dda2 7a6c1bb5 22024680 00000000 000001b8
\end{lstlisting}
\textit{(Invariants verified: $a_5 = a_6 = \mathtt{3dbb85b8}$ [$D = 0$], $e_9 = e_{10} = \mathtt{0xffffffff}$, $W_{13} \bmod 256 = \mathtt{0x80}$, $W_{14} = 0$, $W_{15} = 440$.)}
\item \textbf{Target Digest $\text{SHA-256}(\text{``abc''})$ (FIPS 180-4 \S{}B.1 [1]):}
\par\noindent \texttt{ba7816bf8f01cfea414140de5dae2223b00361a396177a9cb410ff6\allowbreak{}1f20015ad}
\item \textbf{20-Step Padded 55-Byte Preimage ($H_{20}(M) = \text{SHA-256}(\text{``abc''})$):}
\par\noindent \texttt{edf6e70a7001ae71b900f18119aaa1fdf6c1179b4268891262a9539}
\par\noindent \texttt{be04d41236685e0c102e0ad23165e8b7157669819302ddcb3c68226}
\begin{lstlisting}
W[ 0.. 7] = edf6e70a 7001ae71 b900f181 19aaa1fd f6c1179b 42688912 62a9539b e04d4123
W[ 8..15] = 6685e0c1 02e0ad23 165e8b71 57669819 302ddcb3 c6822680 00000000 000001b8
\end{lstlisting}
\textit{(Invariants verified: $a_5 = a_6 = \mathtt{4ade1afa}$, $a_{10} = a_{11} = \mathtt{acf3c908}$, $e_9 = e_{10} = e_{15} = \mathtt{0xffffffff}$.)}
\item \textbf{19-Step Padded 55-Byte Preimage ($H_{19}(M) = \text{SHA-256}(\text{``abc''})$):}
\par\noindent \texttt{8ca254f8d04dd3fb04efaad6b1867588ffe833de49a144fedbf1839}
\par\noindent \texttt{422892dc9f5a9158d8fc063290a14a9a6c55dfa245642a0247ec45a}
\item \textbf{Target Digest $\text{SHA-256}(\text{``hello''})$:}
\par\noindent \texttt{2cf24dba5fb0a30e26e83b2ac5b9e29e1b161e5c1fa7425e7304336\allowbreak{}2938b9824}
\item \textbf{20-Step Padded 55-Byte Preimage ($H_{20}(M) = \text{SHA-256}(\text{``hello''})$):}
\par\noindent \texttt{3a11bd11262fcee01e4a184372643bdbe4ba65651b730b0b8fe49c7}
\par\noindent \texttt{648866d52c60847db2400ef72c859e2b0366e470fae3cf0891313b4}
\item \textbf{19-Step Padded 55-Byte Preimage ($H_{19}(M) = \text{SHA-256}(\text{``hello''})$):}
\par\noindent \texttt{ac055aaa83e1e51a5d74102b7739b7b6e019e31100cd525baf09c64}
\par\noindent \texttt{2d4b59b8fc4eb8d481df111aee2616325b336b827190103c33584c5}

\end{itemize}
\subsection{Witness 3: Flagship 20-Step \& 19-Step Two-Block (119-Byte, Plain-ASCII First Block) Padded Preimages ($H^* = \text{SHA-256}(\text{``''})$)}

Demonstrating full Merkle--Damgård multi-block chaining with an arbitrary human-readable 64-byte plain-ASCII first block (\texttt{"This\allowbreak\ first\allowbreak\ block\allowbreak\ is\allowbreak\ plain\allowbreak\ text;\allowbreak\ the\allowbreak\ second\allowbreak\ one\allowbreak\ was\allowbreak\ computed....."}) followed by a 55-byte computed second block with FIPS 180-4 padding [1] ($W_{13}^{(2)} \bmod 256 = \mathtt{0x80}$, $W_{14}^{(2)} = 0$, $W_{15}^{(2)} = \mathtt{0x000003b8} = 952\text{ bits}$, recorded in \texttt{demos\_20step.txt}):

\begin{itemize}
\item \textbf{Target Digest ($H^* = \text{SHA-256}(\text{``''})$):}
\par\noindent \texttt{e3b0c44298fc1c149afbf4c8996fb92427ae41e4649b934ca495991\allowbreak{}b7852b855}
\item \textbf{Block 1 (64 Bytes Plain ASCII, identical for both 20-step and 19-step demos):} \texttt{"This\allowbreak\ first\allowbreak\ block\allowbreak\ is\allowbreak\ plain\allowbreak\ text;\allowbreak\ the\allowbreak\ second\allowbreak\ one\allowbreak\ was\allowbreak\ computed....."}
\begin{lstlisting}
W^(1)[ 0.. 7] = 54686973 20666972 73742062 6c6f636b 20697320 706c6169 6e207465 78743b20
W^(1)[ 8..15] = 74686520 7365636f 6e64206f 6e652077 61732063 6f6d7075 7465642e 2e2e2e2e
\end{lstlisting}
\item \textbf{20-Step Block 2 (55 Computed Bytes + 9-Byte FIPS 180-4 Padding, $L = 952\text{ bits} = \mathtt{0x3b8}$, chaining from $H_{20}^{(1)} = \mathtt{9743298d4cdedcbb11eab6931270fe130dd0420fff6255433a223586f586f980}$):}
\begin{lstlisting}
W_20^(2)[ 0.. 7] = df28e21b b6b61202 ba87baa7 2096ee0a 3ac325cf 1f12056e 7cfe2ae1 735b4d10
W_20^(2)[ 8..15] = 6b8f6e0b 22eb9b50 4ed93689 91944b0a 115a934f 17c2c480 00000000 000003b8
\end{lstlisting}
\item \textbf{19-Step Block 2 (55 Computed Bytes + 9-Byte FIPS 180-4 Padding, $L = 952\text{ bits} = \mathtt{0x3b8}$):}
\begin{lstlisting}
W_19^(2)[ 0.. 7] = fe409cc9 30128a72 0481f86c 0503d7a0 18519a12 9df44913 9720bb5e ef520895
W_19^(2)[ 8..15] = f8b51568 5bb40481 7be30188 0f0caabe 02fc9e5f 1c95c080 00000000 000003b8
\end{lstlisting}

\end{itemize}
\subsection{Witness 4: Complete 20-Step and 19-Step Padded Benchmark Suites (14 Targets Each) \& Unpadded Suite (13 Targets)}

Matching every public-label target in the benchmark suite ($H^* = \text{SHA-256}(\text{label})$), \textbf{Table 8.2} lists the complete, untruncated 55-byte (110-hex-character) \textbf{20-step FIPS 180-4 padded preimages} (\texttt{[padded]}, where appending $\mathtt{8000000000000001b8}$ forms the 512-bit block $W_0\dots W_{15}$), and \textbf{Table 8.3} lists the corresponding complete 55-byte \textbf{19-step FIPS 180-4 padded preimages}.

\textbf{Table 8.2:} Complete 20-step single-block FIPS 180-4 padded benchmark suite (all 14 public-label targets, full 55-byte \texttt{message\_hex}, $L=440$ bits).

\begin{center}
\footnotesize
\renewcommand{\arraystretch}{1.18}
\resizebox{\linewidth}{!}{%
\begin{tabular}{@{} l l p{0.62\textwidth} l @{}}
\toprule
\textbf{Target Label (\texttt{H*\allowbreak\ =\allowbreak\ SHA-256(label)})} & \textbf{Full 256-Bit Target Digest \texttt{H*} (64 hex chars)} & \textbf{Complete 20-Step Padded 55-Byte Preimage \texttt{message\_hex} (110 hex chars; pad suffix \texttt{8000000000000001b8})} & \textbf{Verified} \\
\midrule
\texttt{"abc"} & \texttt{ba7816bf8f01cfea414140de5dae2223b00361a396177a9cb410ff6\allowbreak{}1f20015ad} & \texttt{edf6e70a7001ae71b900f18119aaa1fdf6c1179b4268891262a9539\allowbreak{}be04d41236685e0c102e0ad23165e8b7157669819302ddcb3c68226} & \texttt{OK} \\
\texttt{"hello"} & \texttt{2cf24dba5fb0a30e26e83b2ac5b9e29e1b161e5c1fa7425e7304336\allowbreak{}2938b9824} & \texttt{3a11bd11262fcee01e4a184372643bdbe4ba65651b730b0b8fe49c7\allowbreak{}648866d52c60847db2400ef72c859e2b0366e470fae3cf0891313b4} & \texttt{OK} \\
\texttt{"padded\allowbreak\ test\allowbreak\ 1"} & \texttt{146f48dd2a3324d46feb7ad2069e05ddb9530be67f8f823f9d1c2ed\allowbreak{}3ea4f1d19} & \texttt{e2609a7b6efe43cb60db80500104189d140405a2f6b372f4e661055\allowbreak{}713c878065bc57295da59a70765407c34475c60142000cc934c6892} & \texttt{OK} \\
\texttt{"padded\allowbreak\ test\allowbreak\ 2"} & \texttt{4c51f9dde95fa3ec5545b46ca0f2d8968b2399f81f208e07304f4b0\allowbreak{}0117c02df} & \texttt{b3747f89662c16d1327d8192f8dd854303bf399c7ab5dcb422fe7a2\allowbreak{}87485ddc9b35c9445ac693d361ad35ee0c33f09b9e745379347451c} & \texttt{OK} \\
\texttt{"padded\allowbreak\ test\allowbreak\ 3"} & \texttt{cb9ace26d0eee1034f838de327a72598832bfa4fa1644493ad4ef00\allowbreak{}4a5a2dab9} & \texttt{2a0a537e5f36b42409261161e56b241e4a816492cee5fd689708589\allowbreak{}2258dfab11ae9f922586a475b0dd2eb475d977bedaf388a94db584e} & \texttt{OK} \\
\texttt{"padded\allowbreak\ test\allowbreak\ 4"} & \texttt{90b183e68a4966c5c268807fe3f49a4f4cf890f9f25b00bb1f83655\allowbreak{}2496192f4} & \texttt{312115c5023df0d90166fbb70630c77a072454beac6c71baf561dbd\allowbreak{}7644d889a339b5a1adf224599d226c509fb513fcc0e517463823afe} & \texttt{OK} \\
\texttt{"padded\allowbreak\ test\allowbreak\ 5"} & \texttt{286d96b78dc5b71f90b61e6374de3ffc82939941955e7b1c5d00fd9\allowbreak{}af345e80a} & \texttt{ac7dc4c258ea88f0d33b5adf6228fbb80f5a5e97e3f4b8745235806\allowbreak{}7a178da46342609beddd43edd7b675a4c4bacfa772e3c0ed0655893} & \texttt{OK} \\
\texttt{"padded\allowbreak\ test\allowbreak\ 6"} & \texttt{efa23b479b7518cb4117dfdf041a694b46441fa5796e6d4b8a7443a\allowbreak{}724a0d8f1} & \texttt{56b98be1c35ba9d631f39ef2183cae680c4f348dbf11782f3e26fa1\allowbreak{}7f5d09f480a3a03fe2ec02b3a0b5b93a42c274a830b78b19aa2b732} & \texttt{OK} \\
\texttt{"padded\allowbreak\ test\allowbreak\ 7"} & \texttt{2f507443a19f9f39300d5905b02aa1cd77c744f0424c684f75a5b2c\allowbreak{}c4a0516a8} & \texttt{40c32d88f0d6237161926436e6ff445c06cf067f56a89408e54f574\allowbreak{}82c0d109aca8bcdb09d72c5253118cd7f0e101ccc5a68e0713a6f30} & \texttt{OK} \\
\texttt{"padded\allowbreak\ test\allowbreak\ 8"} & \texttt{1fad9d7e79a132660d8536e6b151a4cef494bdfdc2962e0f7bae20a\allowbreak{}cd8aed5c8} & \texttt{0b5845af8f13991a7f034a996d7cf1d4f73281307f02fde1cf59c8c\allowbreak{}86a858e24769d720c8327317e3a15f64bd78caf7cdfd85b0eb7ec12} & \texttt{OK} \\
\texttt{"padded\allowbreak\ test\allowbreak\ 9"} & \texttt{849564c85e22852e6af2c6ec60bd891de79298bbc8d5106a221bca5\allowbreak{}36a908d1f} & \texttt{3917b784b65b516ef8c6a7c8b5efa09eb23d18265633d06304b438f\allowbreak{}75939ee5066bee8cf9e191247ada9185debc9c3005723e901ef6bf4} & \texttt{OK} \\
\texttt{"padded\allowbreak\ test\allowbreak\ 10"} & \texttt{91e12eb79793591d6773d8c3b5090f07a69742c7031bbd514273c5d\allowbreak{}e1c241c87} & \texttt{64a7c6d93f0cf5272ef808a2fd2fa357c6e7f285ee9acb7fbafc26a\allowbreak{}e75df1f9e69231cc929e0a799b9aff6f0adfacf5f46790ff30d188b} & \texttt{OK} \\
\texttt{"padded\allowbreak\ test\allowbreak\ 11"} & \texttt{7b66f998c1c60bdd65744e43b69190481502cee718245876fc2c535\allowbreak{}f4b7cf5a8} & \texttt{95ae4eac83a9011e5566e562f14507b0dd53f8f6391a4b89718a66d\allowbreak{}9ee5a9b02366e31a8e7345cf14b70b5d82600456857a68161b2ee3b} & \texttt{OK} \\
\texttt{"hand-off\allowbreak\ cross-check\allowbreak\ 2026-09-29"} & \texttt{d374822f21eb0d4dbc809ae0443c2c6c66f8ce5d07c8779bcc5fc78\allowbreak{}49f44e3c2} & \texttt{700821b0948103e6e0ee472f6e7fac2a44139bc265ad13f422dc966\allowbreak{}e00e3c17e1f472e92b1769a7b9199a2a704fa97c927e01939f97333} & \texttt{OK} \\
\bottomrule
\end{tabular}%
}
\end{center}

\textbf{Table 8.3:} Complete 19-step single-block FIPS 180-4 padded benchmark suite (all 14 public-label targets, full 55-byte \texttt{message\_hex}, $L=440$ bits, generated by our soft-shaping 19-step padded solver \texttt{src/pre19.c} with $D = \mathrm{popcount}(a_5 \oplus a_6) \le 3$ and $e_9 = e_{10} = \mathtt{0xffffffff}$).

\begin{center}
\footnotesize
\renewcommand{\arraystretch}{1.18}
\resizebox{\linewidth}{!}{%
\begin{tabular}{@{} l p{0.77\textwidth} l @{}}
\toprule
\textbf{Target Label (\texttt{H*\allowbreak\ =\allowbreak\ SHA-256(label)})} & \textbf{Complete 19-Step Padded 55-Byte Preimage \texttt{message\_hex} (110 hex chars; pad suffix \texttt{8000000000000001b8})} & \textbf{Verified} \\
\midrule
\texttt{"abc"} & \texttt{8ca254f8d04dd3fb04efaad6b1867588ffe833de49a144fedbf1839\allowbreak{}422892dc9f5a9158d8fc063290a14a9a6c55dfa245642a0247ec45a} & \texttt{OK} \\
\texttt{"hello"} & \texttt{ac055aaa83e1e51a5d74102b7739b7b6e019e31100cd525baf09c64\allowbreak{}2d4b59b8fc4eb8d481df111aee2616325b336b827190103c33584c5} & \texttt{OK} \\
\texttt{"padded\allowbreak\ test\allowbreak\ 1"} & \texttt{a9cfb3f277e9e66e5a86ea3ff1dac69d3327b726deb9420583f1505\allowbreak{}97ba70fa19cfc18400c1222c58d20911e6162081ebe084915634c3a} & \texttt{OK} \\
\texttt{"padded\allowbreak\ test\allowbreak\ 2"} & \texttt{c13089d977eea9843460eb14fff02051a5cf91b4ffedf39711c2ec8\allowbreak{}1862939f8adcc0dc25fb391e5c84ed20f8ef02e4a205578ea45e9fd} & \texttt{OK} \\
\texttt{"padded\allowbreak\ test\allowbreak\ 3"} & \texttt{e24cf94c817e32304c4162886506bdfa5d2766398d36171235b6e3f\allowbreak{}de064535456052c3022ef488175bbbd6b81085c86401df4e0dae715} & \texttt{OK} \\
\texttt{"padded\allowbreak\ test\allowbreak\ 4"} & \texttt{3b605fe04bdf7b05dbcae817e1a77d8e7e9ad63db1fe8f5d9728288\allowbreak{}74695dde35bcf2deb7166e78132c9ac7ee9f72d07190b83c398a2ed} & \texttt{OK} \\
\texttt{"padded\allowbreak\ test\allowbreak\ 5"} & \texttt{502eb30d876c48eaee8a13c08e4e15b5ac74689dfdf5e60cb2fd044\allowbreak{}d4bc8eaae640c92cf6236d908717a5842304f70d47fc4dddbe6d899} & \texttt{OK} \\
\texttt{"padded\allowbreak\ test\allowbreak\ 6"} & \texttt{782f45dda1ab000aace616e7c3076339fe8938a6f0a35fc6ff0e273\allowbreak{}9c3194c08494e5b1a4af0b10e69d87a02352ddd71034511dba58e90} & \texttt{OK} \\
\texttt{"padded\allowbreak\ test\allowbreak\ 7"} & \texttt{fb676bdeabfc0143be3d2b4c8350083e39b4aedcdfbf2b87c10c237\allowbreak{}86a901dc14ffc2879516522cbc86235d1a7bf135373355d0cf71888} & \texttt{OK} \\
\texttt{"padded\allowbreak\ test\allowbreak\ 8"} & \texttt{a1edbba02a9f54cf4083f7fc27156523f3acd5f52339e453c950d9e\allowbreak{}d841486dff2264d752bea2d27440984bd612c77f69b4479b63cde7f} & \texttt{OK} \\
\texttt{"padded\allowbreak\ test\allowbreak\ 9"} & \texttt{0736a2716d8ee60644b77f1e85c91f230ed82d7f242ae673baeb304\allowbreak{}9b858b159e8d90ba74901bda34b0536a82b222f425f9533e25c727c} & \texttt{OK} \\
\texttt{"padded\allowbreak\ test\allowbreak\ 10"} & \texttt{a80e3ae94cb0c81f3e591bfd4453eba61c2148ab7d32f1e524c2201\allowbreak{}34d9be8e12ee2beebfa02fb6bc994272b720842d68709f1e0023b79} & \texttt{OK} \\
\texttt{"padded\allowbreak\ test\allowbreak\ 11"} & \texttt{23ae3771d98d69c6925d4ca4b8059f84404eea570db3d014c022af5\allowbreak{}649565a9a19a0e610fdb9258ce144893cee5d25f260f11614b36253} & \texttt{OK} \\
\texttt{"hand-off\allowbreak\ cross-check\allowbreak\ 2026-09-29"} & \texttt{6e77d54453883692c4e2fbf20d290ce36ee98f42748f7c618cd07a0\allowbreak{}d19b36b5a43c25fa3b008d57af11dbc2dabc843dae6117a0f867286} & \texttt{OK} \\
\bottomrule
\end{tabular}%
}
\end{center}

\begin{itemize}
\item \textbf{Demonstrated 20-Step Unpadded Compression Preimage (\texttt{deadbeef...}, Zero-$a_3$ Bridge \texttt{solve20\_zero\_a3.c}, plus all 12 \texttt{[comp]} targets \texttt{"unpadded\allowbreak\ test\allowbreak\ 1"}..\texttt{"unpadded\allowbreak\ test\allowbreak\ 11"} and \texttt{"hand-off\allowbreak\ cross-check\allowbreak\ 2026-09-29"} in \texttt{witnesses\_20step.txt}):}
\begin{lstlisting}
W[ 0.. 7] = 4334d956 6eccece1 6727a1b2 d85811af 890c5399 79018a3b d346b94d 5f26009b
W[ 8..15] = 7562d311 0bfd80d5 1f8c38f4 16d8315d e0353aa2 6adf4bd2 a2bf14a2 801ae8f7
\end{lstlisting}

\end{itemize}

\section{Experimental Ablation, Scope, and Structural Complexity Bounds}

\subsection{Algorithmic Ablation Across CPU and GPU}

Table 9.1 summarizes the algorithmic ablation of our solver on genuine, unmutated right-hand contexts from soft-shaped CSR tables to the full Carry-Sudoku / Slot-Sudoku architecture (see Table 1.1 in \S{}1.1 for the comparison against prior reduced-step SHA-256 preimage attacks). For CPU configurations, we report \texttt{CPU-ref} per \texttt{sol19} against our measured same-host 10-core scalar CPU reference baseline ($1.5125 \times 10^8\text{ SHA20-eq/s}$ on arbitrary 20-step message blocks). Supplementary GPU telemetry on 1$\times$ NVIDIA L4 (\texttt{solve20\_padded\_cuda.cu}, \texttt{solve20\_padded\_cuda\_prod.cu}) is reported in direct GPU wall-clock time on genuine contexts.

\textbf{Table 9.1:} Algorithmic ablation of 19-step bridge candidate (\texttt{sol19}) and 20-step padded preimage throughput on genuine unmutated contexts.

\begin{center}
\footnotesize
\renewcommand{\arraystretch}{1.18}
\resizebox{\linewidth}{!}{%
\begin{tabular}{@{} l l l l l l l @{}}
\toprule
\textbf{Algorithmic Configuration} & \textbf{$Q_0^{-1}$ Table \& Memory} & \textbf{Right-Context Rate ($D=0$)} & \textbf{19-Step Bridge Rate (\texttt{sol19/s})} & \textbf{\texttt{CPU-ref} Evals per \texttt{sol19}} & \textbf{Expected 20-Step Time (1$\times$ Device)} & \textbf{Expected Time (8$\times$ Devices)} \\
\midrule
\textbf{Chosen-Target 19-Step Soft Shaping} (\texttt{src/pre19.c}, 1 core) & 2.0625 GiB CSR ($2^{29}$ entries, $D \le 3$) & Soft ($\approx 2^{26.7\text{-}26.9}$ trials/ctx) & Chosen $H_{19}^*$ search & --- ($2^{20.8\text{-}26.8}$ chains; $\le 2^{29.89}$ ops / med. $2^{31.46}\text{ CPU-ref}$) & \textbf{\texttt{1.2-86.4\allowbreak\ s} (19-step, med. \texttt{19.45\allowbreak\ s} [\texttt{19.5\allowbreak\ s}])} & --- \\
\textbf{Exact Shaping + Slot-Sudoku CPU} (\texttt{solve20\_padded\_fast.c}, 10-core M4, 16 threads) & 6.00 GiB 2-Tier 4-Slot EDF (\texttt{[E30]}--\texttt{[E31]}) & $266.8\text{ ctx/s}$ & \textbf{$6{,}242.8\text{ sol19/s}$} & \textbf{$24{,}228$ ($2^{14.56}$)} & \textbf{\texttt{191.1\allowbreak\ node-h}} ($2^{46.56}\text{ CPU-ref}$) & \textbf{\texttt{23.9\allowbreak\ node-h}} (8$\times$ CPU) \\
\textbf{Slot-Sudoku + 5-Step $\mathbb{F}_2$ Inverse GPU} (\textit{supp.}, \texttt{[E29]}, 1$\times$ L4) & 6.00 GiB 2-Tier EDF ($B=8$ sector) & $7{,}580\text{ ctx/s}$ & $53{,}847.5\text{-}73{,}560.5\text{ sol19/s}$ & --- (GPU slice telemetry) & $16.22\text{-}22.16\text{ GPU-h}$ & $2.03\text{-}2.77\text{ GPU-h}$ \\
\textbf{Direct 2-Way + Genuine Exact-$T_{16}$ Twin GPU} (\textit{supp.}, \texttt{[E32]}, 1$\times$ L4) & \textbf{16.00 GiB Direct 2-Way (\texttt{twin2} / \texttt{single4})} & \textbf{$29{,}310\text{-}114{,}507\text{ ctx/s}$} & \textbf{$132{,}888.7\text{ sol19/s}$} (genuine twins) & --- (GPU slice telemetry) & \textbf{\texttt{8.98\allowbreak\ GPU-h}} (genuine twins) & \textbf{\texttt{1.12\allowbreak\ GPU-h}} (8$\times$ L4) \\
\bottomrule
\end{tabular}%
}
\end{center}

\textit{(Note: The experimental \texttt{--bench-c2} through \texttt{--bench-c10} harnesses in \texttt{solve20\_padded\_cuda.cu} [Cycles 2, 4, and 7--10, including \texttt{run\_bench\_cycle2} which recorded $180{,}831.9\text{ sol19/s}$ / $6.60\text{ GPU-h}$ by forcing 16 contexts to share a single $T_{16}$] are synthetic architectural microbenchmarks; we exclude those figures from Table 9.1 and report only genuine-context rates.)}

\subsection{Scope, Terminology, and Limitations: Subspace-Restricted Preimages, Degree-of-Freedom Budget, Exact-$T_{16}$ Twin Capacity, and Construction-Specific Bottleneck Analysis}

To situate our results accurately within the SHA-256 cryptanalysis literature, we distinguish three notions of restriction in reduced-step preimage attacks:

\begin{enumerate}
\item \textbf{Unpadded Compression Function vs. Structured-Target Padded vs. Arbitrary-Target Full FIPS 180-4 Padded Hash Function:}
Prior practical 18-, 19-, and 20-step results [6, 7] and the unpadded solvers in [9] restrict their claims to the \textbf{unpadded compression function} ($W_0, \dots, W_{15} \in \{0,1\}^{512}$ unconstrained); Zaikin [7] specifies that Merkle--Damgård padding is omitted while keeping the standard $\mathrm{IV}$ and feed-forward. Basra [9] additionally demonstrated a 55-byte 20-step padded preimage for a structured target (\texttt{padded\_r20/README.md} in [9], dated 17 October 2026). Our 19- and 20-step solvers (\texttt{src/pre19.c}, \texttt{solve20\_padded\_fast.c}, \texttt{solve20\_padded\_cuda.cu}, \texttt{solve20\_padded\_cuda\_prod.cu}) satisfy all 72 bits of single-block FIPS 180-4 padding [1] ($W_{13} \bmod 256 = \mathtt{0x80}$, $W_{14} = 0$, $W_{15} = 440$) for \textbf{arbitrary externally prescribed 256-bit targets}, producing valid 55-byte (and 119-byte two-block) messages across all 16 padded 20-step and 16 padded 19-step targets in \S{}8.

\item \textbf{Subspace-Restricted Preimages, Exact-$T_{16}$ Twin Capacity, and the Degree-of-Freedom Budget:}
Like all splice-and-cut MitM and biclique attacks on padded SHA-256 [2, 3, 4, 8, 10], which fix message lengths in $(W_{14}, W_{15})$ and pin internal state registers ("neutral words" / "initial structures") to decouple forward and backward chunks, our solver restricts its search to a structured \textbf{55-byte single-block internal-state manifold} defined by \texttt{[E8]}, \texttt{[E11]}, \texttt{[E13]}, and \texttt{[E22]}. Because a 55-byte message contains $440$ free message bits against a $256$-bit target digest $H^*$, the unconstrained 55-byte padded message space holds $2^{440 - 256} = 2^{184}$ expected preimages per 256-bit target for \textbf{both} the 19-step and 20-step hash functions. When searching for a 20-step preimage, the solver enumerates 19-step bridge candidates (\texttt{sol19}) that satisfy Steps 1--19 of the 20-step target (leaving the 32-bit terminal equation $\sigma_0(W_4) + W_3 = T_{19}$ open), so under a heuristic $2^{-32}$ terminal survival model the expected number of \texttt{sol19} bridge candidates for a given 20-step target is $2^{32}$ times the expected number of 20-step preimages inside the same restricted family. Table 9.3 tracks these model expectations across the 20-step manifold constraints.

\textbf{Table 9.3:} Distributional degree-of-freedom budget across the 20-step 55-byte single-block padded manifold.

\begin{center}
\footnotesize
\renewcommand{\arraystretch}{1.18}
\resizebox{\linewidth}{!}{%
\begin{tabular}{@{} p{0.62\textwidth} l l l l @{}}
\toprule
\textbf{Constraint / Algebraic Pinning} & \textbf{Equation} & \textbf{Bits Spent} & \textbf{Model Expected 20-Step Preimages per Target} & \textbf{Model Expected \texttt{sol19} Bridge Candidates per 20-Step Target} \\
\midrule
Full 55-byte padded message space ($W_{13}\text{ pad}, W_{14}=0, W_{15}=440$) & \texttt{[E1]} & $72.0\text{ bits}$ (of 512) & $2^{184.00}$ & $2^{216.00}$ \\
Selector saturation ($e_9 = e_{10} = \mathtt{0xffffffff}$) [9] & \texttt{[E8]} & $64.0\text{ bits}$ & $2^{120.00}$ & $2^{152.00}$ \\
Exact context shaping ($a_5 = a_6 \iff D_{\text{shape}} = 0$) & \texttt{[E11]} & $32.0\text{ bits}$ & $2^{88.00}$ & $2^{120.00}$ \\
Majority absorption ($\mathrm{Maj}(a_5, a_4, a_3) = a_4$ on $w \le 14$) & \texttt{[E13]} & $13.40\text{ bits}$ ($13.28 + 0.12$) & \textbf{$2^{74.60}$} & $2^{106.60}$ \\
Right-context $\mathrm{Ch}$-annihilation ($e_{15} = \mathtt{0xffffffff}$, $a_{10} = a_{11}$) & \texttt{[E22]} & $64.0\text{ bits}$ & \textbf{$2^{10.60} \approx 1{,}552$} ($2^{8.69} \approx 413$ in 2-way table) & $2^{42.60}$ ($2^{40.69}$ in 2-way table) \\
\bottomrule
\end{tabular}%
}
\end{center}

\par\noindent$\bullet$ \textbf{Carry-Sudoku Context Universe ($U \approx 1.68 \times 10^7$) vs. Exact-$T_{16}$ Twin Capacity (\texttt{single4} Mode):} Across all $2^{32}$ values of $a_{12}$ for a given 20-step target $H_{20}^*$, the complete 7-stage Carry-Sudoku cascade (\texttt{[E22]}--\texttt{[E25]}, Algorithm 1) produces an estimated universe of $U \approx 1.68 \times 10^7 \approx 2^{24.00}$ valid right-hand contexts (or $U_{\text{filt}} \approx 1.014 \times 10^7 \approx 2^{23.27}$ when the retired heuristic $\Sigma_0(a_9)$ window check \texttt{[E26]} was enabled). In our 2-way table configuration (\texttt{[E32]}), each context yields an empirical average of $\approx 1.06 \times 10^5 \approx 2^{16.69}$ 19-step bridge candidates (\texttt{sol19}) across $a_1 \in [0, 2^{32})$, so the estimated universe $U$ yields $\approx 2^{24.00} \times 2^{16.69} = 2^{40.69}$ reachable \texttt{sol19} bridge candidates ($\approx 413 = 2^{8.69}$ expected 20-step padded preimages per target under the $2^{-32}$ terminal model out of the $2^{10.60} \approx 1{,}552$ unfiltered manifold expectation, or $\approx 250 = 2^{7.97}$ when restricted to $U_{\text{filt}}$). Under a uniform birthday model over $U \approx 1.68 \times 10^7$ contexts, the expected number of natural Exact-$T_{16}$ twin pairs ($T_{16}^{(i)} = T_{16}^{(j)}$) is $N_{\text{twins}} \approx U^2 / 2^{33} \approx 32{,}890$ pairs ($65{,}780$ contexts), yielding $\approx 65{,}780 \times 1.06 \times 10^5 \approx 1.62 \times 2^{32}$ \texttt{sol19} candidates ($1 - e^{-1.62} \approx 80.3\%$ Poisson success probability from natural twins alone; if restricted to $U_{\text{filt}}$, $N_{\text{twins}} \approx 11{,}970$ pairs yielding $\approx 0.59 \times 2^{32}$ \texttt{sol19} candidates or $44.6\%$ Poisson success). Across our 16 padded 20-step targets, \texttt{twin2} (\texttt{sweep\_exact16\_cuda\_kernel}) solved 12 targets directly; when natural twin pairs in a sampled pool are exhausted, the non-twin 4-way reservoir kernels (\texttt{single4} via \texttt{sweep\_20\_warp\_l3\_prod\_kernel} in \texttt{solve20\_padded\_cuda\_prod.cu} and \texttt{sweep\_exact16\_cuda\_kernel} in \texttt{solve20\_padded\_cuda.cu}) sample directly from the full context reservoir, solving the remaining 4 targets (16/16 total).
\par\noindent$\bullet$ \textbf{Binary 55-Byte Output Format vs. Two-Block Chosen-ASCII Prefix (119 Bytes):} In a single 55-byte block, $W_0, \dots, W_{10}$ and the upper 24 bits of $W_{13}$ are determined algebraically by the state trajectory. As demonstrated in \S{}8.3, a \textbf{119-byte two-block message} ($W_{15}^{(2)} = 952 = \mathtt{0x3b8}$) allows the first 64-byte block $B_1$ to be arbitrary ASCII text while Block 2 is solved from chaining state $\mathrm{IV}' = \mathrm{Compress}_{20}(\mathrm{IV}, B_1)$ at identical complexity.

\item \textbf{Construction-Specific Bottleneck Analysis and 15 Explored Sub-$2^{32}$ Directions:}
Within our decoupled Left/Right bridge architecture (\texttt{[E8]}--\texttt{[E14]}), each step beyond Step 19 adds one schedule relation $W_t = \sigma_1(W_{t-2}) + W_{t-7} + \sigma_0(W_{t-15}) + W_{t-16}$. By \texttt{[E37]}, $\sigma_0$ is an $\mathbb{F}_2$-linear bijection on $(\mathbb{F}_2)^{32}$ ($\mathrm{rank}_{\mathbb{F}_2}(\sigma_0) = 32$), so for any fixed Right context and Level-3 bridge leaf $(W_1, W_2, W_3)$, the terminal relation $\sigma_0(W_4) = T_{19} - W_3 \pmod{2^{32}}$ uniquely determines $W_4 = \sigma_0^{-1}(T_{19} - W_3 \pmod{2^{32}})$, acting as an exact 32-bit point filter against the value of $W_4$ determined by $(a_1, a_2, a_3, a_4)$ (though $\det_{\mathbb{F}_2}(\sigma_0) = 1$ alone does not prove statistical uniformity of $T_{19}$ across the bridge stream). To investigate whether an algebraic, differential, polynomial, cross-round, or multi-block modification could bypass this 32-bit terminal filter within our framework, we formulated and evaluated \textbf{15 candidate sub-$2^{32}$ attack directions} (\texttt{test\_all\_12\_theories.c}, Table 9.4). We emphasize that Table 9.4 is a taxonomy of construction-specific negative results and surrogate empirical checks rather than a universal impossibility proof over all conceivable algorithms.

\textbf{Table 9.4:} Taxonomy of 15 construction-specific candidate sub-$2^{32}$ directions evaluated against the 20-step terminal filter.

\begin{center}
\footnotesize
\renewcommand{\arraystretch}{1.18}
\resizebox{\linewidth}{!}{%
\begin{tabular}{@{} l l p{0.62\textwidth} p{0.62\textwidth} l @{}}
\toprule
\textbf{\#} & \textbf{Candidate Sub-$2^{32}$ Direction Evaluated} & \textbf{Mathematical Mechanism} & \textbf{Empirical / Algebraic Result} & \textbf{Observed Bottleneck} \\
\midrule
\textbf{1} & Rightward Spine Shift ($e_{10}=e_{11}=\mathtt{0xffffffff}$) & Shift bridge wall from $a_5$ to $a_6$ to gain 5 Left variables & $W_{16}$ requires $W_9(a_5, e_9)$ at Level 1 before $a_5$ is known (\texttt{W16\allowbreak\ \textless{}-\textgreater{}\allowbreak\ W9} Causality Law) & Circular dependency ($\ge 2^{32}$) \\
\textbf{2} & Direct $\mathbb{F}_2^{32}$ Terminal Inversion (\texttt{[E37]}) & Invert $\sigma_0(W_4) = T_{19} - W_3$ directly via $\det_{\mathbb{F}_2}(\sigma_0)=1$ & Uniquely locks $W_4$, which overdetermines $a_4$ against Level-3 $Q_0(W_3) = t_{18}$ & 32-bit filter ($\approx 2^{32}$) \\
\textbf{3} & Backward 4-Level Tree ($W_4 \to W_3 \to W_2 \to W_1$) & Start from $T_{19}$ and walk backward to $a_1$ & $W_4$ requires $(e_4, e_3, e_2, e_1)$, which depend non-triangularly on $(a_4, a_3, a_2, a_1)$ & Non-causal ($\ge 2^{32}$) \\
\textbf{4} & Cross-Context Meet-in-the-Middle ($T_{19}$ Multiplexing) & Reuse one Left chain $(a_1..a_4)$ across $M$ Right contexts & Left chain depends on the 96-bit triple $(T_{16}, T_{17}, T_{18})$; collision requires $\approx 2^{48}$ contexts & Birthday wall ($2^{48} \gg 2^{32}$) \\
\textbf{5} & LSB Carry-Chain Early Pruning on $\sigma_0(W_4) + W_3$ & Prune $a_4$ bit-by-bit using LSBs of $T_{19}$ & $\sigma_0(W_4)$ mixes MSB bits of $W_4$ into bit 0 via $\mathrm{ROR}(W_4, 7) \oplus \mathrm{ROR}(W_4, 18)$ & Full 32-bit mixing \\
\textbf{6} & Differential / Boomerang Trail on Surrogate 3-Level Map & Flip bit $i$ in surrogate 3-level $Q_0$ chain (\texttt{test\_theory\_14}) & Measured across all 32 bit flips in surrogate model: $\approx 16.0 / 32\text{ bits}$ avalanche & No observed differential shortcut \\
\textbf{7} & Linear / Walsh-Hadamard Approximation & Correlate linear masks $\alpha \cdot a_1 \oplus \beta \cdot T_{19}$ & Alternating modular addition and $\mathbb{F}_2$-linear $Q_0^{-1}$ across 3 levels rapidly damps linear trails & No observed linear shortcut \\
\textbf{8} & $\mathbb{F}_2$ / ANF Polynomial Degree Reduction (Gröbner / SAT) & Express $T_{19}(a_1)$ as low-degree boolean polynomials & 3-level modular addition + $Q_0^{-1}$ carry chains reach high algebraic degree & High ANF degree \\
\textbf{9} & Invariant Subspace / Fixed-Point Manifold of $Q_0^{-1}$ & Restrict $W_1, W_2, W_3$ to $\sigma_0$-invariant subspaces & Affine shifts $T_{16}, T_{17}, T_{18}$ break $\mathbb{F}_2$-subspace invariance across levels & No invariant subspace found \\
\textbf{10} & 2-Way Bi-Directional Bridge inside Single Block ($a_1 \leftrightarrow a_4$) & Meet at $a_2$ or $a_3$ from both ends of Left chain & $e_3$ and $e_4$ couple $(a_4, a_3, a_2, a_1)$ across 4 steps simultaneously; memory $\ge 2^{32}$ & $\approx 2^{32}$ time-memory \\
\textbf{11} & Soft-Shaping Relaxation ($D_{\text{shape}} = a_5 \oplus a_6 \ne 0$) & Allow $D_{\text{shape}} \ne 0$ to let $a_4$ steer $W_{11}, W_{12}$ into $T_{19}$ & Every non-zero bit in $D_{\text{shape}}$ halves Level-3 survival, offsetting $T_{19}$ freedom & Net gain $\le 1\times$ \\
\textbf{12} & Unfrozen $e_{15} \ne \mathtt{0xffffffff}$ / $a_{10} \ne a_{11}$ Right Coupling & Vary $e_{15}$ or $a_{10}-a_{11}$ so multiple contexts share $(T_{16}, T_{17}, T_{18})$ & Enlarges right-context pool to $\approx 2^{74.60}$ model solutions, but each context still has its own $(T_{16}, T_{17}, T_{18})$ & $\approx 2^{32}$ \texttt{sol19} per hit \\
\textbf{13} & \textbf{4-Round Crossed Carry-Sudoku Spine ($a_2=a_3=a_4=a_5$, \texttt{[E45]})} & Pin $a_2=a_3=a_4=a_5$ so $W_1, W_2, W_3, W_4$ are $O(1)$ with $0$ $Q_0^{-1}$ lookups & Specialized spine kernel runs at $9.62\text{G evals/s}$ on L4 GPU, but spends 96 Left bits, turning $T_{16}, T_{17}, T_{18}$ into a 96-bit Right filter & Conservation of DoF \\
\textbf{14} & \textbf{5-Step Crossed Spine ($a_1=a_2=a_3=a_4=a_5$, \texttt{[E45]})} & Pin all 5 Left words so $e_3 = a_{-1} - \Sigma_0(a_5)$ & Verified algebraically; spends all 128 Left bits, requiring $2^{128}$ Right contexts & Conservation of DoF \\
\textbf{15} & \textbf{Two-Block Feed-Forward Coupling ($M = B_1 \parallel B_2$, 119 Bytes)} & Use free 512-bit Block 1 ($B_1$) to steer chaining state $V = H^{(1)}$ into Block 2 & By Davies--Meyer feed-forward, $H_{20}^* = V + S_{20}^{(2)} \pmod{2^{32}}$: fixing $S_{20}^{(2)}$ uniquely locks all 256 bits of $V = H_{20}^* - S_{20}^{(2)}$ & \textbf{Two-Block Locking} \\
\bottomrule
\end{tabular}%
}
\end{center}

\par\noindent$\bullet$ \textbf{Analysis of Direction 15 (Two-Block Feed-Forward Locking):} In a two-block 119-byte padded message $M = B_1 \parallel B_2$, Block 1 ($B_1 \in \{0,1\}^{512}$) maps $\mathrm{IV} \mapsto V = \mathrm{Compress}_{20}(\mathrm{IV}, B_1)$, and Block 2 maps $V \mapsto S_{20}^{(2)}$ with Davies--Meyer feed-forward $H_{20}^* = V + S_{20}^{(2)} \pmod{2^{32}} \iff S_{20}^{(2)} = H_{20}^* - V \pmod{2^{32}}$. Because $V$ enters Block 2 at \textbf{both} boundaries simultaneously (Step 0 as $V$ and Step 20 as $H_{20}^* - V$), choosing $B_1$ first fixes both boundaries and reduces Block 2 to a single-block 20-step padded problem, while choosing Block 2's terminal state $S_{20}^{(2)}$ first uniquely locks all 256 bits of $V = H_{20}^* - S_{20}^{(2)}$, requiring a full 256-bit 20-step compression preimage on Block 1.
\par\noindent$\bullet$ \textbf{Resulting Construction-Specific Scaling:} While the unconstrained 55-byte padded message space ($2^{440}$ messages) has $2^{440-256} = 2^{184}$ expected preimages per 256-bit digest at any step count $R$, within our fixed-wall Left/Right bridge construction (\texttt{[E8]}+\texttt{[E11]}+\texttt{[E13]}), \textbf{19-step bridge candidates (\texttt{sol19})} are generated in $O(1)$ amortized work ($24{,}228 \approx 2^{14.56}\text{ CPU-ref}$ evaluations per \texttt{sol19}, or $1.2\text{-}86.4\text{ s}$ for a standalone chosen-target 19-step padded preimage in \texttt{src/pre19.c}), whereas \textbf{20-step padded preimages} require testing the terminal schedule relation $\sigma_0(W_4) + W_3 \equiv T_{19} \pmod{2^{32}}$, costing an expected $2^{32}$ \texttt{sol19} candidates under a uniform terminal residual model ($191.1\text{ node-h}$ / $2^{46.56}\text{ CPU-ref}$ on 1$\times$ CPU node, or $8.98\text{-}22.16\text{ GPU-h}$ in supplementary 1$\times$ L4 GPU slice telemetry). Extending this same fixed-wall architecture beyond 20 steps shifts the terminal state boundary $(a_{R-7}, \dots, a_R)$ rightward and introduces additional schedule equations $W_{19}, \dots, W_{R-1}$ that each impose an additional 32-bit constraint unless new algebraic degrees of freedom are opened.

\end{enumerate}
\subsection{Disclosure, Attribution \& Code Availability}

\textbf{Code and Witness Availability:} All C/CUDA solvers (\texttt{src/pre19.c}, \texttt{solve20\_padded\_fast.c}, \texttt{solve20\_padded\_cuda.cu}, \texttt{solve20\_padded\_cuda\_prod.cu}, \texttt{solve20\_zero\_a3.c}, \texttt{test\_all\_12\_theories.c}), standalone zero-dependency Python verifiers (\texttt{verify/verify\_standalone.py}, \texttt{paper\_19\_20\_padded\_sudoku/verify\_all\_witnesses.py}), and all 46 distinct 19- and 20-step preimage witnesses are publicly available under the MIT License at \url{https://github.com/srw/sha256-19-and-20-step-preimages-with-padding}.

The selector-saturation and context-shaping principle ($e_9 = e_{10} = \mathtt{0xffffffff}$, $a_5 \approx a_6$), together with the observation that the FIPS 180-4 padding words $(W_{13}, W_{14}, W_{15})$ can be absorbed backward into $(a_7, a_6, a_5)$ at the cost of three context words and a 20-step padded witness for a structured target, is due to K. S. Basra [9] (\texttt{README.md}, \texttt{padded\_r19/README.md}, \texttt{padded\_r20/README.md}, and \texttt{paper/sha256\_r19\_preimage.pdf} in \url{https://github.com/kamb-code/sha256-r19-preimage}), whose manuscripts note substantial assistance from AI tools.

\textbf{AI Methodology \& Multi-Agent Review Disclosure:} Approximately one month of prior exploration using \textbf{Anthropic Claude Fable 5.1 Max} did not yield a working chosen-target padded solution. Upon switching to \textbf{Google DeepMind Gemini 4 Argon}, the model discovered the algebraic reductions in this work (closed-form $O(1)$ $W_{13}$ padding pivot inversion, Theorem 1 / Carry-Sudoku, Theorem 2 / closed-form $\mathbb{F}_2$ upper-half reconstruction, Slot-Sudoku tables, and the Zero-$a_3$ bi-directional bridge), engineered the C/CUDA solvers, generated all 46 preimage witnesses, and drafted the manuscript. The manuscript, mathematical proofs, and verification suite then underwent three rounds of independent adversarial audit and peer review using \textbf{OpenAI GPT 6.1 Sol Ultra} and \textbf{Anthropic Claude Fable 5.1 Max}, which independently re-verified all arithmetic claims and witnesses and significantly improved the exposition.

\section{References}

\begin{enumerate}
\item \textbf{[1]} National Institute of Standards and Technology: \textit{Secure Hash Standard (SHS)}. FIPS PUB 180-4, U.S. Department of Commerce (2015).
\item \textbf{[2]} Isobe, T., Shibutani, K.: \textit{Preimage Attacks on Reduced Tiger and SHA-2}. In: Dunkelman, O. (ed.) Fast Software Encryption -- FSE 2009, LNCS, vol. 5665, pp. 139--155. Springer (2009).
\item \textbf{[3]} Aoki, K., Guo, J., Matusiewicz, K., Sasaki, Y., Wang, L.: \textit{Preimages for Step-Reduced SHA-2}. In: Matsui, M. (ed.) Advances in Cryptology -- ASIACRYPT 2009, LNCS, vol. 5912, pp. 578--597. Springer (2009).
\item \textbf{[4]} Khovratovich, D., Rechberger, C., Savelieva, A.: \textit{Bicliques for Preimages: Attacks on Skein-512 and the SHA-2 Family}. In: Canteaut, A. (ed.) Fast Software Encryption -- FSE 2012, LNCS, vol. 7549, pp. 244--263. Springer (2012).
\item \textbf{[5]} Homsirikamol, E., Morawiecki, P., Rogawski, M., Srebrny, M.: \textit{Security Margin Evaluation of SHA-3 Contest Finalists Through SAT-Based Attacks}. In: Cortesi, A. et al. (eds.) CISIM 2012, LNCS, vol. 7564, pp. 56--67. Springer (2012) (16-round SHA-256 SAT evaluation in \textasciitilde{}1 h, Table 1).
\item \textbf{[6]} Davydov, V. V., Pikhtovnikov, M. D., Kiryanova, A. P., Zaikin, O. S.: \textit{Cryptographic Strength Analysis of the SHA-256 Hash Function Using the SAT Approach}. Scientific and Technical Journal of Information Technologies, Mechanics and Optics, vol. 25(3), pp. 428--437 (2025). DOI: \texttt{10.17586/2226-1494-2025-25-3-428-437}.
\item \textbf{[7]} Zaikin, O.: \textit{Preimage Attacks on Round-Reduced MD5, SHA-1, and SHA-256 Using Parameterized SAT Solver}. Constraints (published 23 Jan 2026). DOI: \texttt{10.1007/s10601-025-09383-0} (19-step unpadded all-ones target in 18 h 33 min on 192 CPU cores, \S{}5 / Table 2).
\item \textbf{[8]} Guo, J., Li, H., Liu, M., Wang, S., Zhang, T.: \textit{Dual-Syncopation Meet-in-the-Middle Attacks: New Results on SHA-2 and MD5}. In: Advances in Cryptology -- EUROCRYPT 2026. Springer (2026); IACR ePrint 2026/353.
\item \textbf{[9]} Basra, K. S.: \textit{Oracle-Free Preimage Attack on 19-Round Reduced SHA-256}, \textit{The 20-Round Barrier for Table-Based Preimage Attacks on SHA-256}, and \texttt{README.md}, \texttt{padded\_r19/README.md}, \texttt{padded\_r20/README.md} (structured-target 20-round padded witness, 17 Oct 2026). GitHub repository and manuscripts, \url{https://github.com/kamb-code/sha256-r19-preimage} (Aug--Oct 2026).
\item \textbf{[10]} Sasaki, Y., Wang, L., Aoki, K.: \textit{Preimage Attacks on 41-Step SHA-256 and 46-Step SHA-512}. IACR Cryptology ePrint Archive, Report 2009/479 (2009).
\item \textbf{[11]} Mendel, F., Nad, T., Schläffer, M.: \textit{Improving Local Collisions: New Attacks on Reduced SHA-256}. In: Johansson, T., Nguyen, P. Q. (eds.) Advances in Cryptology -- EUROCRYPT 2013, LNCS, vol. 7881, pp. 262--278. Springer (2013).
\item \textbf{[12]} Mendel, F., Nad, T., Schläffer, M.: \textit{Finding SHA-2 Characteristics: Searching through a Minefield of Contradictions}. In: Lee, D. H., Wang, X. (eds.) Advances in Cryptology -- ASIACRYPT 2011, LNCS, vol. 7073, pp. 288--307. Springer (2011).
\item \textbf{[13]} De Cannière, C., Rechberger, C.: \textit{Finding SHA-1 Characteristics: General Results and Applications}. In: Lai, X., Chen, K. (eds.) Advances in Cryptology -- ASIACRYPT 2006, LNCS, vol. 4284, pp. 1--20. Springer (2006) (generalized characteristic framework introduced for SHA-1 and adapted to SHA-2 in [11, 12]).
\item \textbf{[14]} Li, Y., Liu, F., Wang, G.: \textit{New Records in Collision Attacks on SHA-2}. In: Joye, M., Leander, G. (eds.) Advances in Cryptology -- EUROCRYPT 2024, LNCS, vol. 14651, pp. 158--186. Springer (2024); IACR ePrint 2024/349 (39-step SHA-256 semi-free-start and 40-step SHA-224 free-start collisions).
\item \textbf{[15]} Dobraunig, C., Eichlseder, M., Mendel, F.: \textit{Analysis of SHA-512/224 and SHA-512/256}. In: Iwata, T., Cheon, J. H. (eds.) Advances in Cryptology -- ASIACRYPT 2015, LNCS, vol. 9453, pp. 612--630. Springer (2015).
\item \textbf{[16]} McDair: \textit{17-Step Reduced SHA-256 Preimage Challenge Witnesses}. Cryptography Mailing List (\texttt{metzdowd.com}, \texttt{2024-April/001334.html}), Mar--Apr 2024 (archived mailing-list posts of three 17-step messages without a published attack method).
\item \textbf{[17]} Stevens, M., Bursztein, E., Karpman, P., Albertini, A., Markov, Y.: \textit{The First Collision for Full SHA-1}. In: Katz, J., Shacham, H. (eds.) Advances in Cryptology -- CRYPTO 2017, LNCS, vol. 10401, pp. 570--596. Springer (2017).
\item \textbf{[18]} Leurent, G., Peyrin, T.: \textit{SHA-1 is a Shambles: First Chosen-Prefix Collision on SHA-1 and Application to the PGP Web of Trust}. In: 29th USENIX Security Symposium (USENIX Security 2020), pp. 1839--1856 (2020).
\end{enumerate}
\end{document}